# Some things special about NEAs: Geometric and environmental effects on the optical signatures of hydration

S. Potin[1], P.Beck[1,2], B. Schmitt[1], F. Moynier[3]

[1]Université Grenoble Alpes, CNRS, Institut de Planétologie et d'Astrophysique de Grenoble (IPAG), France (414 rue de la Piscine, 38400 Saint-Martin d'Hères, France)

[2]Institut Universitaire de France, Paris, France

[3]Institut de Physique du Globe de Paris (IPGP) (1 Rue Jussieu, 75005 Paris, France)

Corresponding author: sandra.potin@univ-grenoble-alpes.fr

**ABSTRACT**

Here were report on a laboratory study aiming to reproduce specificities of near-Earth Asteroid. We study how the elevated surface temperature, their surface roughness (rock or regolith), as well as observation geometry can affect the absorption features detected on asteroids. For that purpose, we selected a recent carbonaceous chondrite fall, the Mukundpura CM2 chondrite which fell in India in June 2017. Bidirectional reflectance spectroscopy was performed to analyze the effect of the geometrical configuration (incidence, emergence and azimuth angle) on the measurement. Our results show that reflectance spectra obtained under warm environment (NEA-like) tends to show shallower absorption bands compared to low-temperature conditions (MBA-like), but still detectable in our experiments under laboratory timescales. Irreversible alteration of the sample because of the warm environment (from room temperature to 250°C) has been detected as an increase of the spectral slope and a decrease of the band depths (at 0.7μm, 0.9μm and 2.7μm). Comparing the meteoritic chip and the powdered sample, we found that surface texture strongly affects the shape of the reflectance spectra of meteorites and thus of asteroids, where a dust-covered surface presents deeper absorption features. We found that all spectral parameters, such as the reflectance value, spectral slope and possible absorption bands are affected by the geometry of measurement. We observed the disappearance of the 0.7 μm absorption feature at phase angle larger than 120°, but the 3μm band remains detectable on all measured spectra.

1. INTRODUCTION

Many meteorite groups record the action of water at some point of their geological history (Brearley 2006). Fingerprints of aqueous alteration have been identified in the form of secondary minerals such as carbonates, phyllosilicates or iron oxides. Asteroids, which are the astronomical counterpart of meteorites, also display evidence of aqueous alteration. A first evidence, derived from visible ground-based optical observations, is the presence of absorption bands at 0.7 and 0.9 μm, which can be attributed to Fe bearing phyllosilicates (Vilas 2008). A more direct evidence of the presence of hydrated or hydoxylated minerals at the surface of asteroids is the presence of an absorption band around 3-μm. For main-belt asteroids (MBAs), observations in this spectral region are only possible for the brightest objects (Vilas 1994; Takir and Emery 2012). Spectral surveys have shown that objects with an absorption band at 0.7 μm always show the presence of a 3-μm feature, but that the reciprocal is not true (Usui et al. 2019; Vilas 1994).

Two space missions are presently orbiting possibly volatile-rich near-Earth asteroids (NEAs): Hayabusa-2 (Kitazato et al. 2019) and OSIRIS-Rex (Hamilton et al. 2019). While hydration seems to be a general process among the MBAs population (Fornasier et al. 2014), evidence of hydrated near-Earth asteroid is scarce (Rivkin et al. 2002). Rivkin and DeMeo (2018) calculated the theoretical amount of hydrated NEAs according to their spectral group and found 17 ± 3% for the Ch-group and 43 ± 6% corresponding to C-complex bodies. However, observations from Stuart and Binzel (2004) and Carry et al. (2016) resulted in the statistics of 6 ± 3% of NEAs showing signs of Ch asteroids, and 16 ± 7% showing signatures of C-complex asteroids. While NEAs are certainly more accessible for sample-return purposes, it is not known today whether they experienced specific, or more intense, geological processes when compared to main-belt asteroids. First, it has been shown that the orbital evolution of the NEA (101955) Bennu, the target of the OSIRIS-REx mission, could have led to a significant heating of the surface (over 500 K (Delbo and Michel 2011)) by absorption of solar radiation, which is much higher than the surface temperature experienced in the main-belt (Schorghofer 2008). Second, many NEAs seem to exhibit a rubble pile structure. These rubble-pile asteroids are interpreted as re-agglomerated fragments of a prior object that was catastrophically destroyed following a hypervelocity collision (Michel, Benz, and Richardson 2003, 2004). Last, small NEAs visited by space missions appear devoid of a significant amount of dust-like regolith (Fujiwara et al. 2006), which is at odd with larger objects visited by space missions (Veverka et al. 1997).

In this article, we investigate how the environment of NEAs and the observations themselves can influence the signatures of aqueous alteration. For that purpose, we selected a freshly fallen aqueously altered meteorite, the Mukundpura CM2 chondrite. The impact of thermal processing on the infrared signature has been studied in situ using a high-temperature, high-vacuum reflectance cell. The effect of observation geometry and the impact of surface texture (comparison between rock and dust no larger than a few hundreds of microns) have been investigated using a custom-made reflectance-goniometer (Potin et al. 2018).

2. **Sample and methods**

   **2.1 Meteorite sample**

The Mukundpura meteorite fell in India on June 6, 2017 (Tripathi, Dixit, and Bhandari 2018) and was immediately collected by the authorities, reducing the risk of post-fall alteration. The first petrographic observations of the newly fallen meteorite were performed using a Zeiss polarizing microscope and electron microprobe (Ray and Shukla 2018)and characterized the meteorite as belonging to the CM group.

Mukundpura presents a dark matrix and several millimeter-size bright inclusions (see Figure **1**). The sample used in this study was taken from the exterior of the whole meteorite and thus showed a smooth fusion crust on one of its side. Only the interior of the sample, the side opposite to the crust, was studied in reflectance spectroscopy.

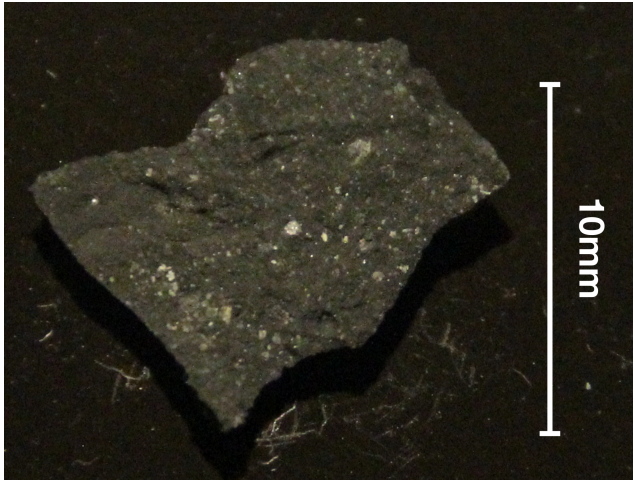

Figure 1: Picture of the Mukundpura meteorite sample used in this study (interior side showing).

This sample was first studied as a chip of the meteorite, then powdered to investigate the effect of texture on the reflectance spectra.

## 2.2 Bidirectional reflectance spectroscopy

### 2.2.1 Definition and methods for bidirectional reflectance spectroscopy under fixed geometry, room temperature and atmospheric pressure.

Reflectance spectroscopy studies are conducted using the custom-made bidirectional reflectance goniometers SHINE (SpectropHotometer with variable INcidence and Emergence) (Brissaud et al. 2004) and SHADOWS (Spectrophotometer with cHanging Angles for the Detection Of Weak Signals) (Potin et al. 2018), whose methods of measurement are quite similar. The goniometer illuminates the sample with a monochromatic light between 340 nm and 5000 nm and under a specific direction

(incidence angle). Table 1 summarizes the different spectral resolutions over the whole spectral range.

**Table 1: Spectral resolution of the reflectance spectrum**

| Spectral range | Spectral resolution |
|---|---|
| 400 nm – 679 nm | 4.85 nm – 4.75 nm |
| 680 nm – 1499 nm | 9.71 nm – 9.38 nm |
| 1500 nm – 2999 nm | 19.42 nm – 18.73 nm |
| 3000 nm – 5000 nm | 38.84 nm – 38.44 nm |

The reflected light is measured under fixed values of emergence and azimuth angles. All three parameters, incidence, emergence and azimuth angles, constitute the geometrical configuration under which the spectra will be recorded. Figure 2 presents a schematic view of the reflectance measurement and the different angles involved.

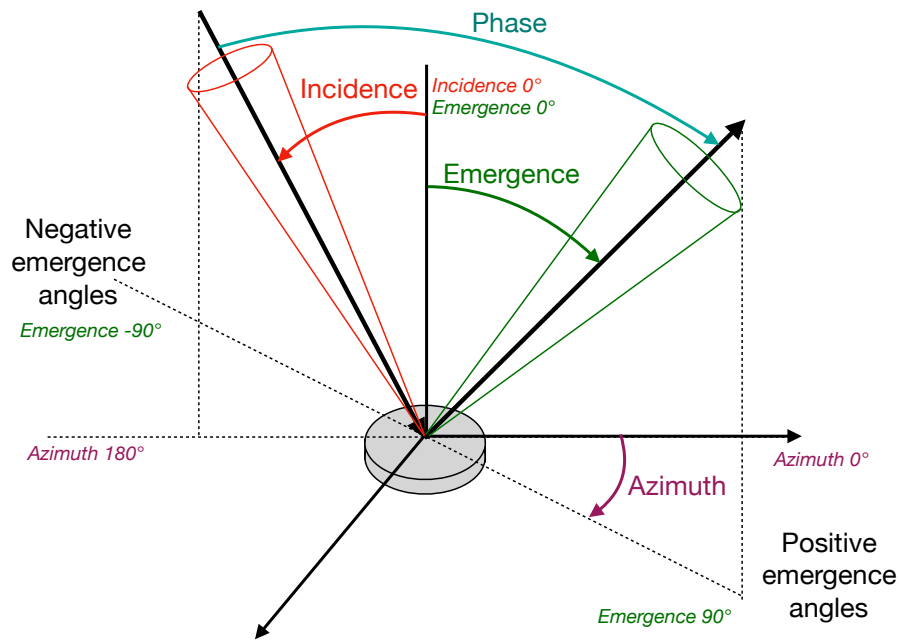

**Figure 2: Schematic definition of the geometrical configuration, incidence, emergence and azimuth angles, for the reflectance spectroscopy. Picture from Potin et al. (2018).**

Nominal configuration for a "single spectrum" reflectance spectroscopy study consists of a nadir illumination (incidence angle of 0°) and an emergence angle of 30° in the principal plane (azimuth 0°). The angular resolutions are ±2.9° for the incidence angle and ±2.05° for the emergence and azimuth angles.

**2.2.2 Reflectance spectroscopy under asteroidal conditions.**

Low-temperature reflectance spectra are acquired using SHINE coupled with the double environmental chamber CarboN-IR (Beck et al. 2015; Grisolle et al. 2014). The outer cell is put under vacuum while gaseous nitrogen is injected inside the inner cell to increase the thermal exchanges and ensure thermalization of the sample. This technique reduces the thermal gradient inside the meteorite but needs two sapphire windows to close the outer and inner cells. The light passing through these two windows creates reflections between the sample and the inner cell window, but also between the windows themselves. The small absorption of the sapphire added to the multiples reflections reduce the contrast of reflectance measured on a sample. To evaluate the effect of the two windows on the measurement, reflectance spectra of a calibration target were acquired, with and without the sapphire windows. Using these spectra, the effects of the windows have been removed from the meteorite data. The spectra were acquired, with a nadir illumination and emergence angle of 30°, from 80K to room temperature 290K every 10K, and from 400 nm to 1000 nm with a 10 nm spectral sampling.

Reflectance spectroscopy at high temperature can be conducted with SHADOWS using a heating vacuum cell, nicknamed MIRAGE (Mesures en InfraRouge sous Atmosphère Gazeuse et Etuvée) (see Figure 3).

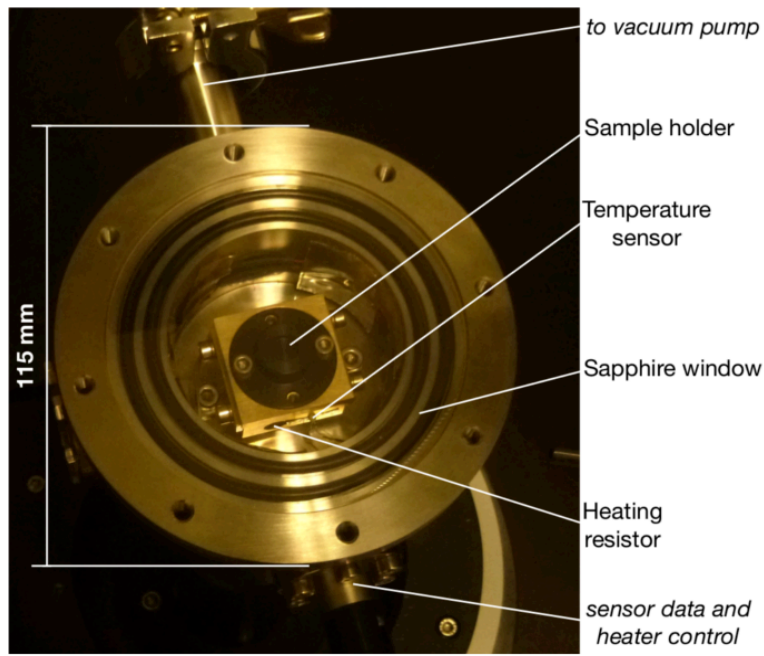

**Figure 3: Annotated picture of the MIRAGE cell.**

This cell only needs one sapphire window, and the presence of reflections between the window and the sample has already been investigated and corrected through the work on the SERAC cell (Spectroscopie En Réflexion sous Atmosphère Contrôlée) (Pommerol et al. 2009). Other sources of error need to be taken into account with MIRAGE, but were previously unknown on SERAC, due to the difference between the illumination methods of the two setups. To remove all these effects, reflectance spectra of Mukundpura were acquired with and without the sapphire window, and used to correct the instrumental biases due to the cell on all the acquisitions. The spectra were acquired from 360 nm to 1200 nm every 10 nm in order to analyze the shape of the 700 nm and 900 nm bands. Acquisitions are set from room temperature (294 K) to 510K every 20K during the increase of temperature, and from 490K to room temperature every 40K when the cell is cooling down. The illumination is fixed at nadir, and the observation at 30°.

### 2.2.3 Bidirectional Reflectance Distribution Function (BRDF)

To simulate observations from a spacecraft orbiting its target or from a ground-based observatory, one has to take into account the scattering geometry of the system: the angle of the light illuminating the surface, as well as the angle from which the spectra are acquired. In the laboratory, bidirectional reflectance spectra were acquired with SHADOWS for 70 different geometrical configurations: incidence, emergence and azimuth angles. The incidence angle was set from nadir to grazing illumination (0° to 60° with a 20° step), and for each incidence, spectra were acquired at several measurement angles from -70° to 70°, corresponding to grazing emergence respectively in the backscattering and forward scattering directions in the principal plane, with a 10° step. To analyze the lateral scattering of the sample, spectra at incidence 20° with the same list of emergence angles were taken with a 30° azimuth angle outside the principal plane. The SHADOWS goniometer cannot measure light reflected at a phase angle lower than 5° so opposition spectra are absent from the dataset. The spectral resolution is identical to the resolutions displayed in Table 1. The angular resolutions are ±2.9° for the incidence angle and ±2.05° for the emergence and azimuth angles.

The series of spectra has been first acquired on the chip. The studied surface originated from a broken piece of a larger sample and therefore was not horizontal when placed on the goniometer (tilt of about 5°). This was taken into account during the analysis and photometric corrections were applied to the data. The same set of reflectance spectra have been acquired after manual grinding of the meteorite, to compare the photometric and spectral variations due to surface texture.

## 3. Reflectance spectroscopy

### 3.1 Spectral features of the Mukundpura CM chondrite

The reflectance spectrum of the powdered meteorite is displayed on Figure 4. A first series of spectra was acquired under vacuum, starting at room temperature then at 60°C. There was no evidence of degassing of the sample.

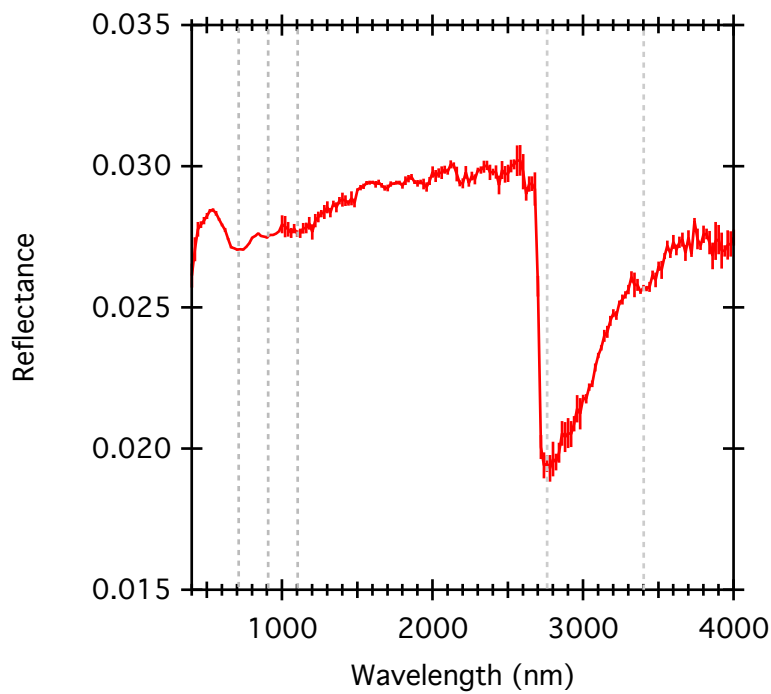

**Figure 4: Reflectance spectrum of the Mukundpura meteorite (powder) with a nadir illumination and an emergence angle of 30°. The grey doted lines mark the position of the typical CM absorption bands at 700 nm, 900 nm, 1100 nm, 2760 nm and 3400 nm.**

The reflectance spectrum shows all characteristics of CM chondrites (Cloutis, Hudon, et al. 2011): (1) a low reflectance value, less than 3% over the whole spectral range, (2) the $Fe^{3+}$-$Fe^{2+}$ charge-transfer band at 700 nm (Tomeoka, Mcsween, and Buseck 1989),

(3) the $Fe^{2+}$ crystal field transition bands at 900 and 1100 nm, (4) organics features at 3400 nm and (5) the –OH stretching band at 2750 nm related to the presence of Mg-OH bearing phyllosilicates.

The beginning of a strong UV absorption feature is visible shortwards of 500 nm, though it is not possible to analyze this band with the current data starting at 400 nm. This absorption feature is likely due to metal-O charge transfer (Cloutis et al. 2008).

The detection of both the iron features (700 nm and 900 nm) and the Mg-rich phyllosilicates band at 2750 nm indicates an extensive aqueous alteration of the meteorite on the parent body (Takir et al. 2013; Beck et al. 2010).

### 3.2 Assessing the impact of regolithisation: comparison between the chip and the powdered sample

Reflectance spectroscopy of the meteoritic chip was conducted under the same geometrical configuration as the powder. The sample was placed on its fusion crust on the sample holder so that the studied surface was revealing the interior of the meteorite. Both spectra are presented in Figure 5.

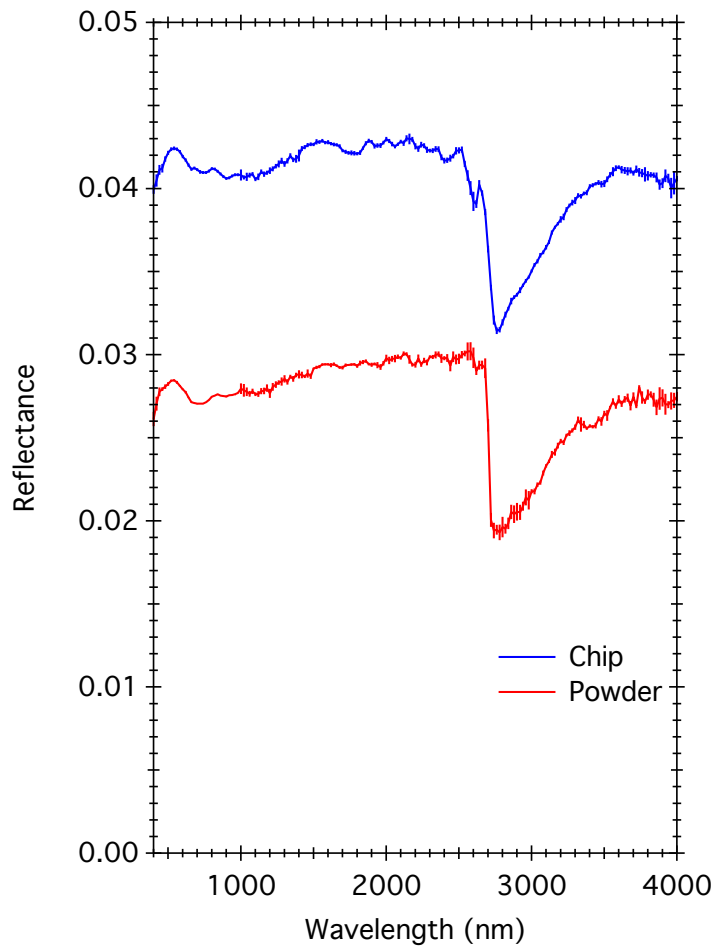

**Figure 5: Reflectance spectra of the raw meteorite (blue) and after manual powdering (red).**

The major difference between the two sets of spectra is the reflectance value of the continuum, around 4.2% in the case of the raw meteorite and around 2.8% for the powder. This behavior is distinct from the "standard" behavior observed for pure material when the grain size is decreased, which is to observe an increase in reflectance with decreasing grain size (Pommerol and Schmitt 2008b). Several possible explanations emerge. The first explanation is related to petrography: the surface of the raw meteorite is heterogeneous due to the presence of mm-sized chondrules and bright inclusions, while the powder has all the elements finely mixed into a rather homogeneous volume. Opaques "hidden" beneath the surface will be more visible in the powder spectra, lowering the global amount of reflected light. The model developed by Hapke (1981) states that the reflectance of an intimate mixtures depends on the viewing geometry and properties of the particles in the mixtures such as single scattering albedo, diameters, porosity and mass fraction. The spectra of mixtures composed of materials with contrasted absorption properties will depend on the style of the mixture. In the case of a spatial mixture of two spectral endmembers, the resulting spectrum is a linear mix of the endmember spectra, weighted by their surface area. The spectra of the raw meteorite, where the size of the bright and dark patches are less than 1 mm² can be interpreted by a linear mixture of the bright and dark endmembers. When powdering the meteorite, the style of mixture between the dark and bright materials will change, since it will become an intimate mixture. In that case, mixtures are expected to be linear in single scattering albedo space, and therefore the mixture of the spectra becomes highly non-linear. For the same amount of bright and dark material, the intimate mixture (powder) will be darker than in the case of a geographic mixture (chip) (Pommerol and Schmitt 2008b).

The second explanation could be the difference of porosity between the two samples. Theory and laboratory measurements agree that increasing the porosity of a sample lowers the general reflectance (Hapke 2008; Cloutis et al. 2018; Shepard and Helfenstein 2011; Cloutis, Hiroi, et al. 2011) The presence of bright inclusions coupled with the differences in both grain size and porosity between the samples can explain the higher reflectance of the raw meteorite.

Both samples present the same absorption features, but with different band depth and broadness. Relative band depths are calculated considering a linear continuum between the two inflexion points of the bands and using the following formula:

$$BD = 1 - \frac{R^c_{band}}{R^c_{continuum}}$$

where c is the central wavelength of the absorption band, $R^c_{band}$ and $R^c_{continuum}$ are respectively the measured reflectance of the sample and the interpolated reflectance of the continuum at wavelength c. The depth of the UV absorption band is calculated as the ratio between the reflectance measured at 540 nm and 400 nm:

$$UV\ ratio = \frac{R^{540nm}}{R^{400\,m}}$$

Table 2 presents the calculated band depths of the UV absorption, the $Fe^{2+}$ 700 nm band, the phyllosilicates 3 µm feature and the organics band around 3400 nm.

**Table 2: Band depths of the Vis-NIR absorption features and UV ratio calculated for the chip and powdered samples.**

|  | UV ratio | 700 nm band | 3 µm band | Organics |
|---|---|---|---|---|
| Chip | 1.0618 ± 0.0074 | 0.0139 ± 0.0037 | 0.1992 ± 0.0090 | 0.0080 ± 0.0061 |
| Powdered meteorite | 1.089 ± 0.018 | 0.0306 ± 0.0017 | 0.3252 ± 0.0009 | 0.0281 ± 0.0088 |

We found that grinding of the sample affects the absorption features, resulting in a deeppening of the bands. Again, the size of the grains will affect the depth of the absorption features, following the general trend where band depth increases with increasing grain size. It should be noted that all absorption bands are detectable in both reflectance spectra. This implies that even in the absence of a well developed fine-grained regolith, as seems to be the case for some near-earth objects (Benner et al. 2008; Hasegawa et al. 2008), the water-related features should be detectable. In order words, the lack of dust-size regolith does not explain the scarcity of water-related absorption signatures among the NEAs population.

## 4 Effects of temperature on the optical absorptions

Several processes may warm the surface of small bodies, NEAs included. First, the initial decay of extinct radionuclide such as $^{26}$Al, could have been strong enough to heat up and differentiate small bodies (Grimm and Mcsween 1993,1989; Mahan et al. 2018), but this process was not efficient for km-size objects (Ghosh et al. 2006). Secondly, impacts can generate enough heat to locally metamorphose or even melt the surface (Bland et al. 2014), though the models cannot describe a global increase of the temperature of the object. This is consistent with some of the heated meteorites showing foliation or shock-induced deformations (Nakamura 2006). The initial accretion process is based on low energy impacts and thus can be disregarded as a source of heating for the small bodies (Nakamura 2005). Then, electromagnetic induction due to solar winds during the early T Tauri phase can induce enough heat to alter the mineralogy of asteroids. But as the solar winds usually occur on the poles of the stars, they are unlikely to have produced enough heat in the mid-plane where the small bodies are concentrated (Mahan et al. 2018). Last but not least, the relatively close distance between the Sun and the NEAs over their lifetime induces an elevated sunlight flux on the surface, and so warming up the small body. This process depends on albedo, surface roughness and thermal inertia. An elevated thermal inertia will reduce the difference of temperature between the direct sub-solar surface (strongly heated) and the rest of the body (rather cool) and the rotation period, as well as the pole position, of the asteroid will also have an impact on the temperature at the surface. In the case of a slow-rotating body or an asteroid oriented so that its rotation axis is near the orbital plane with a pole facing toward the Sun, the sub-solar point will accumulate a lot of heat and will reach higher temperatures than the night-side, and the whole surface of the body will present a wide range of temperatures that strongly vary with time. On the contrary, fast-rotating asteroids, or with their rotation axis almost perpendicular to the orbital plane, will present a rather homogeneous and moderate temperature on their surface. For example, according to models, (101955) Bennu could have undergone temperatures as high as 500K (Delbo and Michel 2011), while other NEAs could have experienced temperatures over 1500K during a close approach to the Sun (Marchi et al. 2009). Note however that Dunn et al. (2013) calculated that the temperature of 72 NEAs is distributed between 143 and 300K.

Reflectance spectra of asteroids can be affected by surface temperature, an effect which was studied for S-type asteroids and ordinary chondrites, but which is much less known for C-type asteroids. In this section, we present the spectral variations of the powdered Mukundpura meteorite under a wide range of environmental conditions, from cryogenic to high temperature. We focus on the iron features in the visible range.

To characterize the spectral variations across changes in temperature, the reflectance spectra are adjusted using a linear continuum and three modified Gaussian profiles (Sunshine, Pieters, and Pratt 1990) representing the UV, 700 nm and 900 nm absorption bands. The modified Gaussian model (MGM) includes a correction factor acting on the symmetry of the left and right wings of the profile. Its value can be fixed empirically. The least residual errors were achieved with a correction factor of 1.12. For each MGM, so for each absorption feature, the amplitude, band position and full width at half maximum (FWHM) are set as free parameters. The resulting model is as follow:

$$R(\lambda) = A\lambda + B + MGM(\lambda, a_{UV}, p_{UV}, \sigma_{UV}) + MGM(\lambda, a_{700nm}, p_{700nm}, \sigma_{700nm}) + MGM(\lambda, a_{900nm}, p_{900nm}, \sigma_{900nm})$$

with R(λ) the calculated reflectance at the wavelength λ, A and B respectively the slope and offset of the continuum, and $a_X, p_X$ and $\sigma_X$ the amplitude, position and FWHM (linked to σ) of the band X. The model parameters for the UV absorption may differ from reality, as the data cover less than half the band.

### 4.1 Variations from cryogenic to high temperatures.

In order to highlight the effect of temperature on the sample, Figure 6 presents a comparison between the extreme spectra, obtained at 80 K and 510 K, and the acquisition at room temperature.

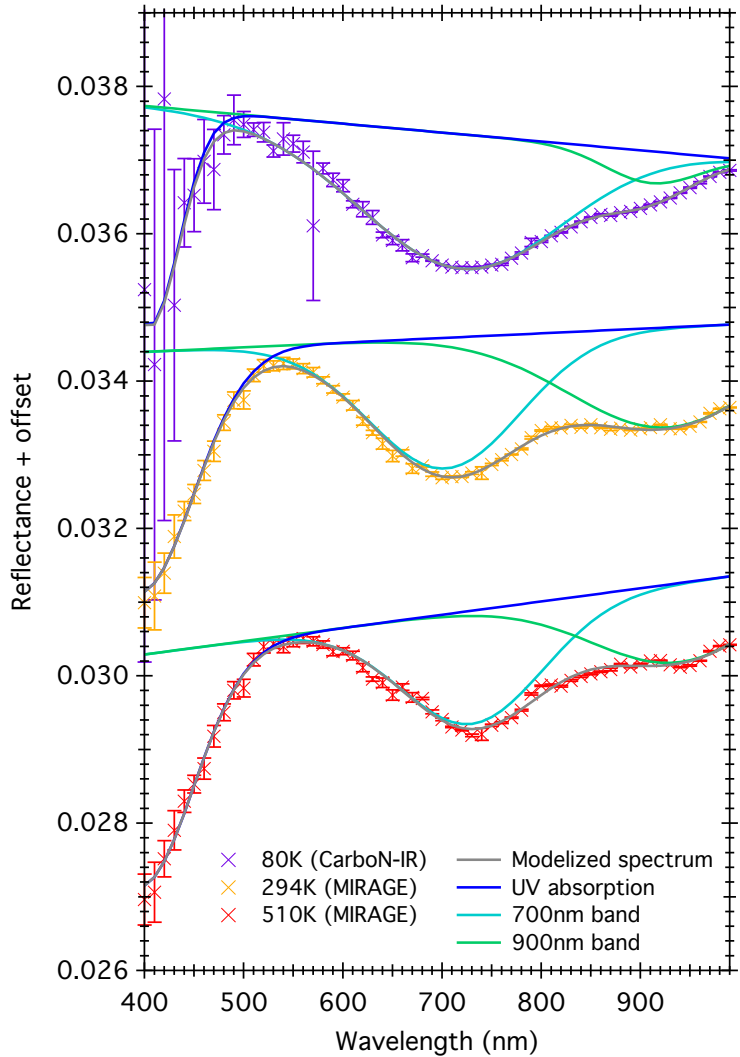

**Figure 6: Comparison between the reflectance spectra of the powdered Mukundpura meteorite acquired at 80K (purple), room temperature (orange) and 510K (red). The modeled spectra (grey) and the different modeled absorption bands, UV (navy blue), 700 nm (blue), and 900nm (green) are also presented. An offset of +0.003 and -0.003 has been added respectively to the 80K and 510K spectra.**

In Figure 6, subtle variations of the spectra related to temperature can be observed. The fit parameters for each spectrum are compared in Table 3.

|                          | 80K           | 294K          | 510K          |
| ------------------------ | ------------- | ------------- | ------------- |
| Continuum slope (nm$^{-1}$) | -1.210.10$^{-6}$ | 6.299.10$^{-7}$ | 1.797.10$^{-6}$ |
| Continuum offset         | 0.0352        | 0.0341        | 0.0326        |
| BD$_{700nm}$ (%)         | 0.183         | 0.177         | 0.153         |
| P$_{700nm}$ (nm)         | 710           | 702           | 730           |
| FWHM$_{700\ nm}$ (nm)    | 330           | 252           | 239           |
| BD$_{900nm}$ (%)         | 0.043         | 0.136         | 0.107         |
| P$_{900nm}$ (nm)         | 910           | 926           | 940           |
| FWHM$_{900\ nm}$ (nm)    | 167           | 327           | 298           |
| BD$_{UV}$ (%)            | 0.298         | 0.327         | 0.314         |
| P$_{UV}$ (nm)            | 404           | 394           | 396           |
| FWHM$_{UV}$ (nm)         | 105           | 176           | 171           |
| Reflectance max          | 0.0344        | 0.0342        | 0.0334        |
| Position of the max (nm) | 491           | 539           | 560           |

**Table 3: Parameters of the modeled spectra at each temperature.**

The increase of temperature induces a modification of the continuum slope, as seen by the spectrum at 80K having a negative (blue) slope, while the spectra acquired at room temperature and 510K present a positive (red) slope. A decrease of reflectance of the continuum is observed at 510K, representing 7.4% relative to the initial value at 80K. Variations of the absorption features as well as the continuum induce an apparent displacement and decrease of the maximum of reflectance, from 0.0344 of reflectance at 491 nm in the spectrum acquired at 80K, to a reflectance of 0.0334 at 560 nm for the spectrum at 510K. The reflectance maximum also decreases with increasing temperature, losing 0.001 of reflectance between the two extreme spectra.

In the case of the low temperature measurements, a blue shift of nearly 20 nm of the position of the 700 nm band can be observed when compared to the high temperature measurement. The band gets broader and deeper, as the FWHM gains around 97 nm and the depth increases by a factor 1.2.  It is interesting to note that opposite effects occur in the case of the 900 nm band, except for the position: with decreasing temperature, the band is shifted by 30 nm towards shorter wavelengths, the band depth is reduced by a factor 2.5 and its FWHM is reduced by approximately 130nm.

Fe-related absorptions can be expected to be sensitive to temperature (Hinrichs and Lucey 2002). The 700 nm feature in CM chondrites is interpreted by $Fe^{2+}$-$Fe^{3+}$ charge transfer ($Fe^{2+}+Fe^{3+}$ ->$Fe^{3+}+Fe^{2+}$) within a phyllosilicate structure. Charge transfer bands in –OH bearing silicates generally display an increase of their intensity under low-temperature, as  observed by Smith (1977). As measured in transmission (absorbance), the intensity of the $Fe^{2+}$-$Fe^{3+}$ charge transfer band of chlorite and biotite has been measured to increase by a factor of 2 when cooling the sample to liquid helium temperature (5.5 K), and by about 25 % when cooling the sample to 80K. This effect was interpreted by a depopulation of one or more electronic levels close to the ground state of the $Fe^{2+}+Fe^{3}$ ->$Fe^{3+}+Fe^{2+}$ transition (Smith 1977). While our lowest temperature measurements are not as low as liquid helium, they are cold enough to induce a significant increase in the depth of the 0.7 μm feature. Note that a broadening of the 700nm band was also observed by Smith (1977) study, which could explain the change in the position of the visible maxima from around 539 nm (room temperature) to 491 nm (80 K), when added to a possible displacement and broadening of the UV absorption.

## 4.2 Irreversible alteration at high temperatures

Reflectance spectra of Mukundpura were recorded during a high temperature experiment. Spectra were acquired during a heating and cooling cycle of the sample and are presented in Figure 7.

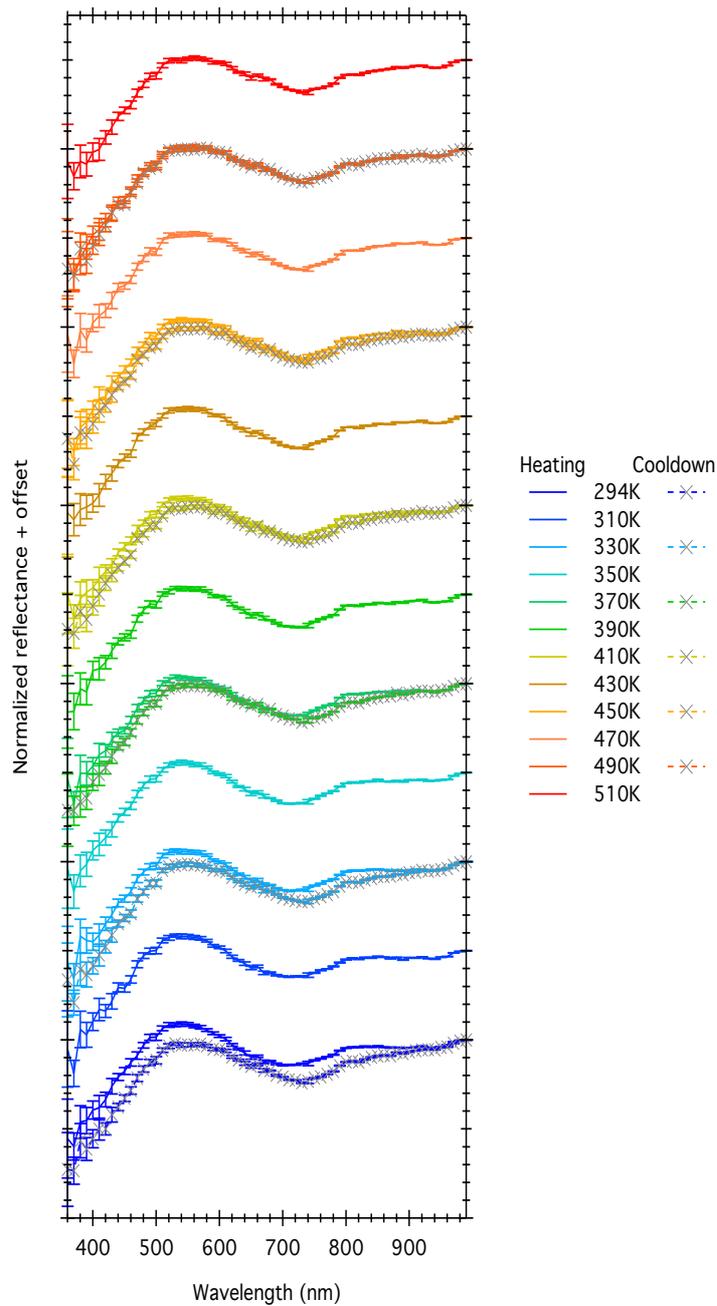

**Figure 7: Normalized reflectance spectra of the powdered Mukundpura meteorite acquired during the increase of temperature from 294K to 510K (solid lines) and during cooling (dashed lined and grey markers). The different offsets have been added for clarity but are identical for spectra acquired at the same temperature.**

The high temperature experiments reveal irreversible alterations of the sample, as seen when comparing the two spectra taken at 294K, at the beginning and at the end of the experiment. The oxidation of the sample by ambient air is not relevant in our measurements as they were performed under secondary vacuum.

The difference between the two spectra is a combination of variations in the continuum of the spectrum and in the 3 detectable absorptions bands. The differences involve an apparent reddening of the sample as the spectrum acquired after heating presents a stronger spectral slope compared to the initial measurement, an apparent displacement of the 700 nm band to longer wavelengths, and both a shift and decrease of the reflectance maximum around 550nm. The whole series of spectra are adjusted using the model described above. Resulting modeled spectra of the first and last measurements of the experiment are presented in Figure 8.

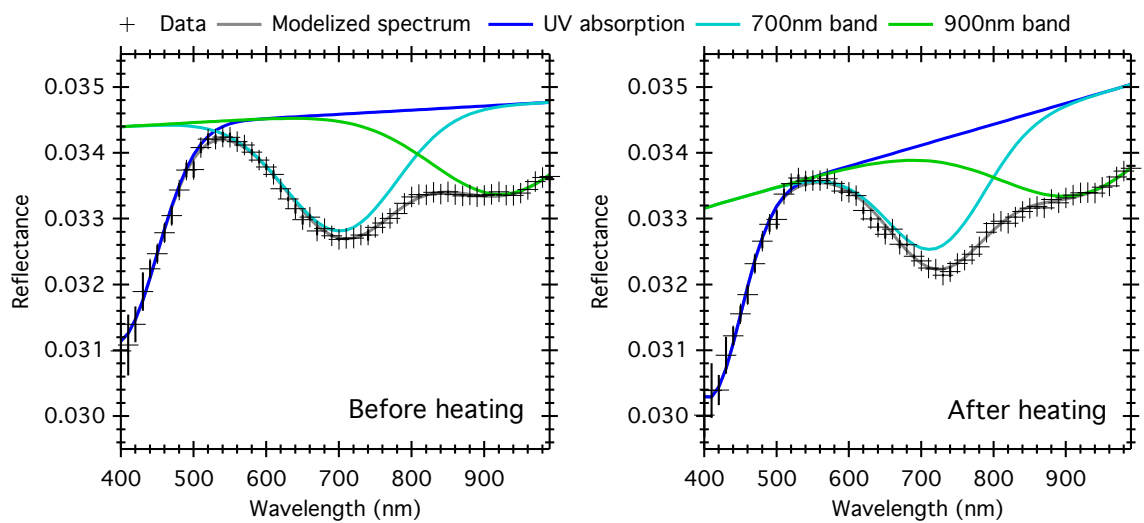

**Figure 8: Reflectance spectra of powdered Mukundpura meteorite at room temperature (294 K) before (left) and after (right) warming to 510 K, as well as the MGM fit to the data (Navy blue: continuum and UV absorption. Blue: continuum and 700nm band, Green: continuum and 900nm band)**

The variation the model parameters calculated for the spectra measured during the heating and cooling cycle, is presented in Figure 9.

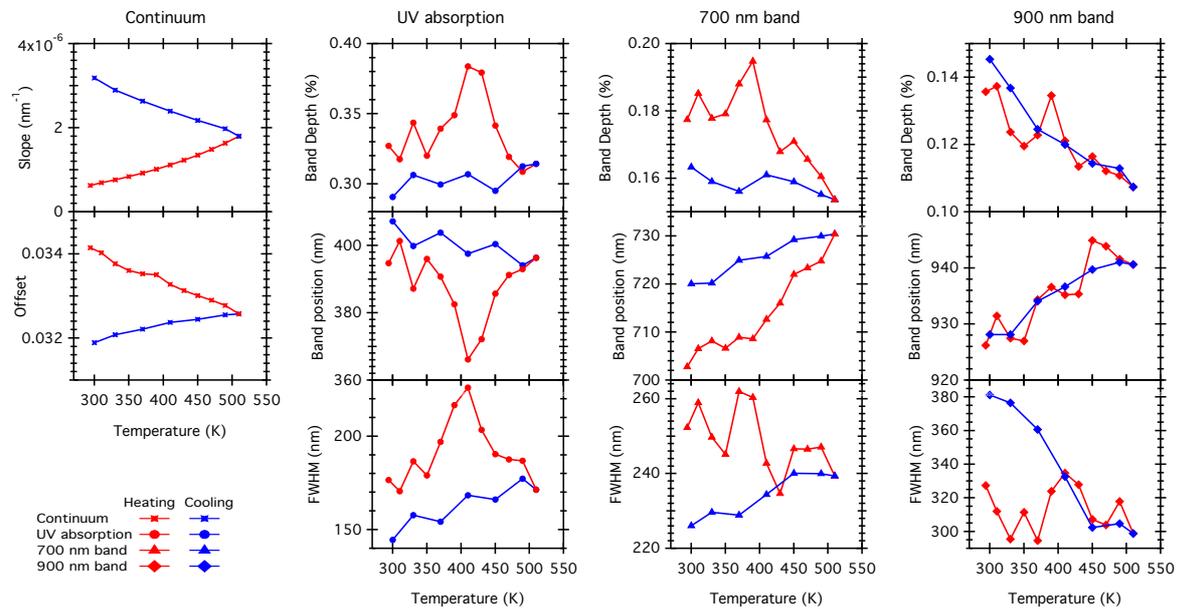

**Figure 9: Variations of the continuum and absorption bands during the heating (red) and cooling (blue) cycle of the sample.**

Based on this analysis, the continuum slope and all the three absorption features show an irreversible alteration due to temperature. In the case of Mukundpura, the temperature cycle up to 510 K and down leads to an increase of the continuum slope, from $6.30 \cdot 10^{-7}$ nm$^{-1}$ to $3.18 \cdot 10^{-6}$ nm$^{-1}$ after the thermal ramp, coupled with a decrease of the general reflectance level represented by the continuum offset, from 0.0341 to 0.0319.

With increasing temperature, the 700 nm band first increases and then decreases in amplitude, while getting narrower and shifting towards longer wavelengths. During cooling, the amplitude of this feature only marginally changes and returns to a value only slightly lower than prior heating (0.163% against 0.177%). Upon heating, the position of the 700 nm band changes significantly. When heating to 520 K the maxima of absorption shifts from 703 to 730 nm, and upon cooling it returns to a value of 720 nm. The position of the 700 nm band is thus not reversible upon heating. This effect could be interpreted by the occurrence of redox process ($Fe^{3+}$ reduction of $Fe^{2+}$ oxidation) within the phylosilicates occurring upon heating, since the oxygen fugacity is sensitive to temperature.

The 900 nm also reveals some variability with temperature but in a different way than the 700 nm band. The amplitude of the band decreases upon heating from 0.130% to 0.107% while the position shifts to higher wavelength by about 15 nm (from 926 to 941 nm). Unlike the 700 nm band, these changes are reversible when returning to ambient temperature (except the FWHM which significantly increases upon cooling). The 900 nm band is interpreted as a crystal field band, which are known to shift, broaden and change in intensity with temperature (Burns 1993), in a reversible manner.

The combined spectral variations due to temperature induce the apparent displacement of the maximum of reflectance from 530 nm to 558 nm at the end of the thermal cycle. The value of the maximum reflectance has been altered as well, slightly decreasing from 0.0342 to 0.0335 after the experiment.

Irreversible alterations also occurred outside the visible spectral range. To compare the different effects observed, the same thermal cycle was imposed on 2 other CM chondrites. Figure 10 presents the reflectance spectra over the whole spectral range of the 3 samples before and after the experiment. As the spectroscopy measurements on the whole spectral range was conducted after the temperature ramp, the samples were prepared into another sample holder, the surfaces have slightly changed between the two experiments. To remove this photometry effect, the spectra are normalized at 560nm.

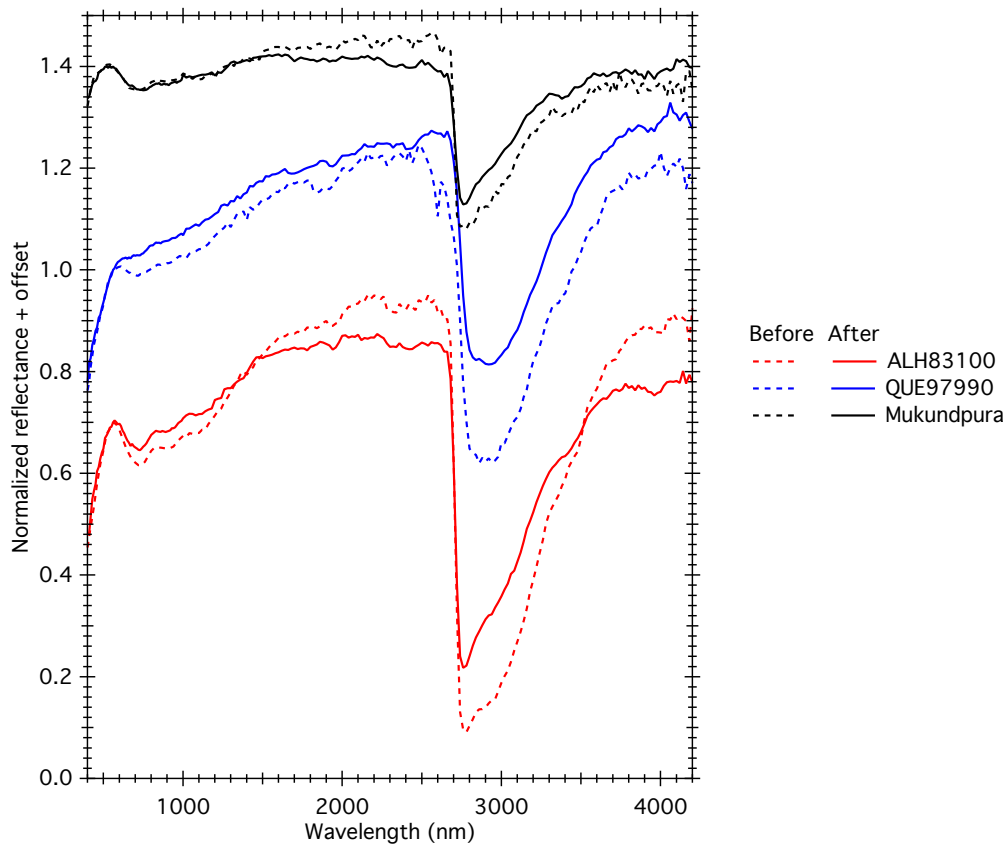

**Figure 10: Normalized reflectance spectra of the CM chondrites ALH83100 (red), QUE97990 (blue) and Mukundpura (black) acquired before (doted line) and after (solid line) a thermal cycle at 520 K. The spectra are normalized at 560 nm and offseted for clarity, -0.3 and +0.4 respectively for ALH83100 and Mukundpura. All samples were manually powdered before the experiment.**

The thermal experiment resulted in another alteration, this time of the 3μm absorption band. Table 4 presents the band depth and width of the 3μm band before and after the thermal cycle for the 3 chondrites.

**Table 4: Band depth and FWHM of the 3µm band before and after the thermal cycle for the 3 CM chondrites.**

|  | Band depth (%) | | FWHM (nm) | |
|---|---|---|---|---|
|  | Before | After | Before | After |
| Mukundpura | 32.52 ± 0.08 | 26.17 ± 0.30 | 410 | 370 |
| ALH83100 | 66.86 ± 0.39 | 54.27 ± 0.17 | 560 | 490 |
| QUE97990 | 49.43 ± 0.57 | 35.21 ± 0.15 | 650 | 530 |

For each sample, the thermal heating reduces both the depth and width of the band. A faint displacement of the band is also detected, which can be explained by the faster reduction of one of the component of the band compared to the others. The alteration of the 3µm band can be due to the release of the absorbed water molecules, and the water contained in (oxy)hydroxide minerals (Garenne et al. 2014).

The present analysis of the reflectance spectra obtained under NEA-like temperatures for CM chondrites reveals that, at laboratory timescale, the presence and so the detection of 700 nm, 900 nm and 3µm bands is not affected by temperature, though the bands can be strongly weakened. Still, a shift of the 700 nm band and a narrowing of the 3µm feature can be predicted for objects that experienced significant thermal processing. Conversely, the shape and mostly the position of the 700nm ands 3µm features could potentially be used as a criterion of primitiveness of Ch and Cgh asteroids.

## 5  Bidirectional reflectance spectroscopy: effects of observation geometry

During a flyby or when orbiting around an asteroid, the observation geometry is very variable (incidence, observation and azimuth angles), and differ from the observation geometry under which laboratory data are typically obtained. This geometry depends on the incidence angle of the sunlight on the surface, and the position of the spacecraft. Figure 11 presents the resulting spectra of the powdered Mukundpura meteorite. The same set of spectra was also acquired on the meteoritic chip.

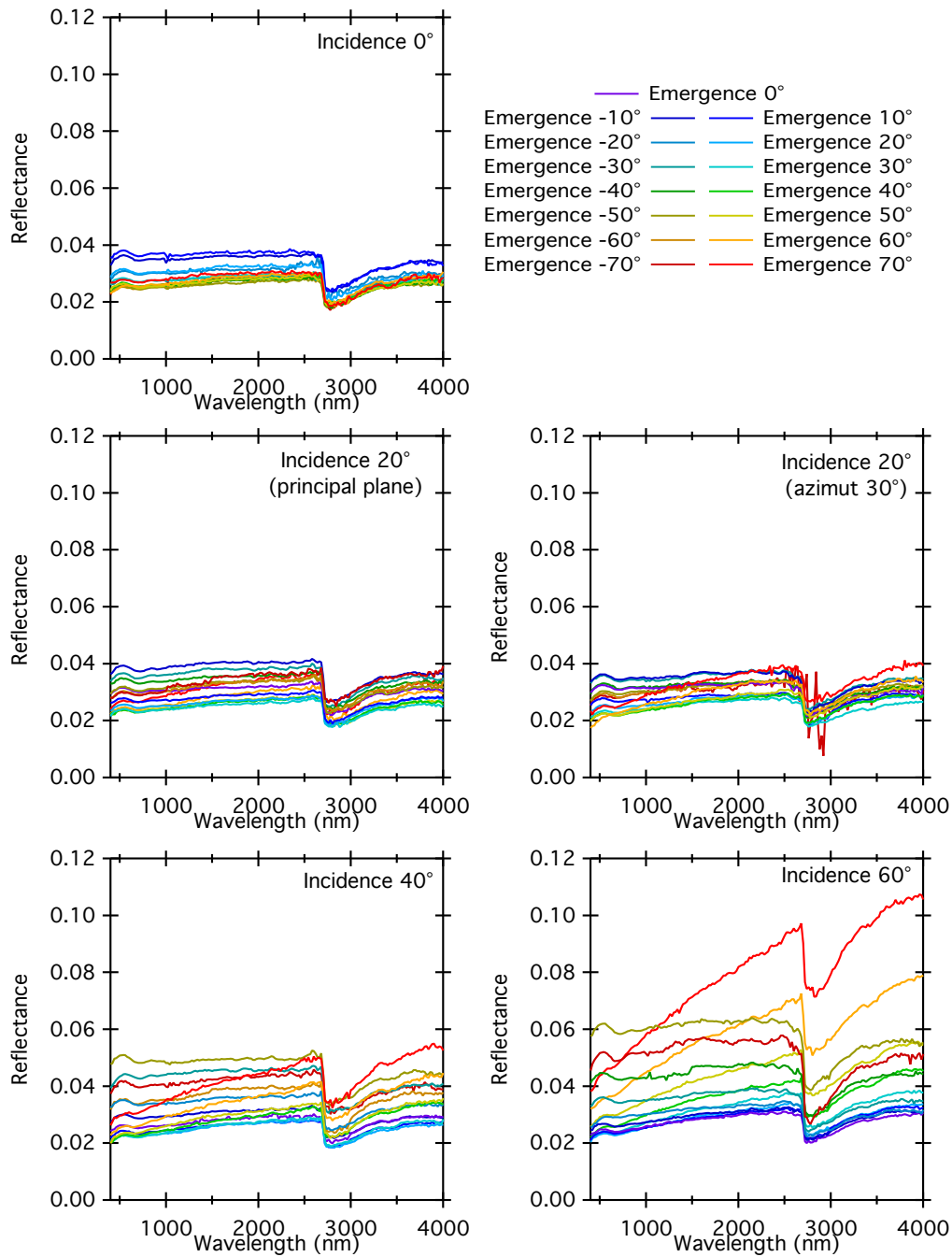

**Figure 11:** Complete set of spectra of the powdered Mukundpura meteorite acquired in bidirectional reflectance distribution function mode. Each panel corresponds to an incidence angle: 0° (top panel), 20° in the principal plane (middle left panel) and at 30° azimuth (middle right panel), 40° (bottom left panel) and 60° (bottom right panel).

The dependence of the reflectance spectra to the geometrical configuration is easily seen on Figure 11. Spectra acquired with an incidence angle of 0° present the same slope, and the photometric level stays between 2 and 4% of reflectance, whatever the emergence angle. But spectra acquired under an incidence angle of 60° show drastic variations according to emergence angle. The reflectance value at 4 µm spreads from 2 to 11%, and an increase of the spectral slope is observed with increasing phase angle, an effect called phase reddening.

In this section, we investigate the different variations of the reflectance spectra when the geometrical configuration is changed. As the two samples, the chip and powder have been analyzed under the same set of geometrical configurations, variations due to the texture of the sample can also be analyzed.

### 5.1 Photometric dependence with the geometry

The bidirectional reflectance distribution function (BRDF) presented on Figure **12** displays the measured reflectance values at different emergence angles for several incidence angles. The polar representation of a BRDF enables a quick overview of the distribution of the reflected light (see Figure **12**). In this representation, the BRDF of a perfectly lambertian surface would appear as a half circle.

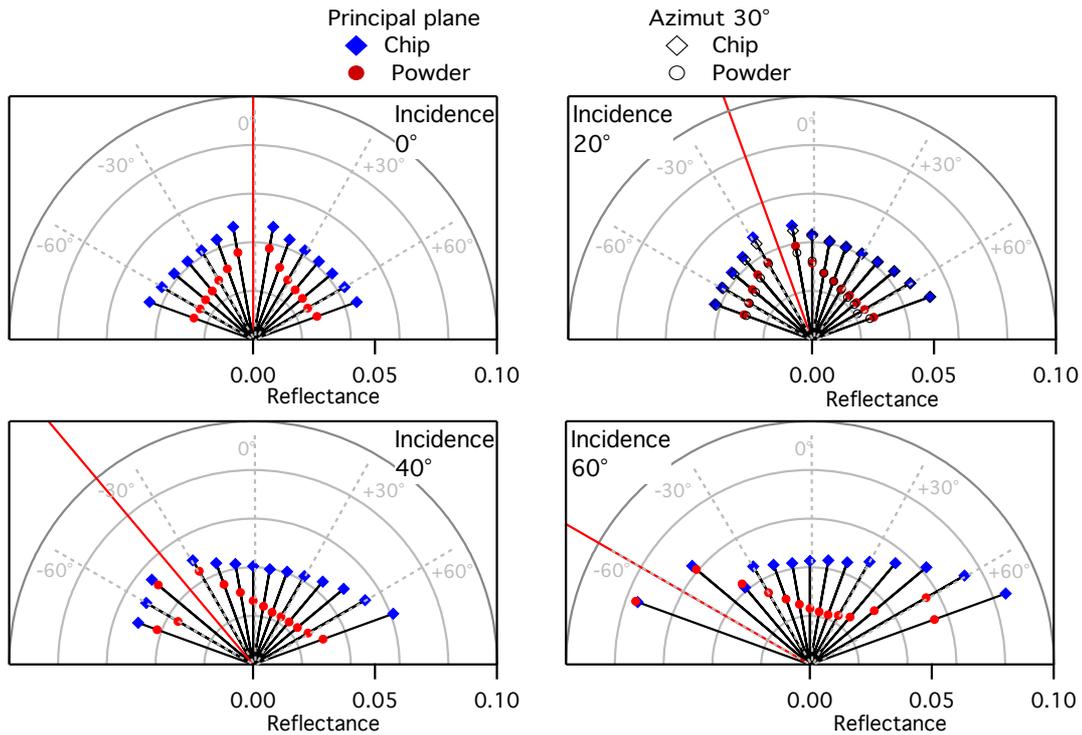

Figure 12: Bidirectional reflectance distribution function of Mukundpura at 560nm. The red line represents the incidence angle and the value of reflectance at each emergence angle is displayed by the colored dots (Blue diamonds: meteoritic chip in the principal plane. Red circles: powder in the principal plane. Open squares: raw meteorite at 30° azimuth angle and 20° incidence angle. Open circles: powder at 30° azimuth angle and 20° incidence angle).

For all geometrical configurations investigated, the chip presents a higher reflectance than the powder. The causes of this effect have been already discussed in 3.2.

The BRDFs shows an increase in reflectivity around the incidence direction (backscattering) and towards high phase angles (forward scattering). It is important to note that both samples show these two behaviors, being more marked when the incidence is close to grazing (60°) and more pronounced on the powder.

Irregularities in the shape, composition, and refractive index of the particle will broaden the forward and backward scattering (Hapke 2002).

This increase of reflectance with decreasing phase angle (backscattering) is due to a combination of two effects. First, the shadows hiding opposition effect (SHOE) where, at phase angle close to 0°, the shadow of one particle over the others around is the smallest, resulting in a strong increase of reflectance in the incidence direction (Hapke 1986). The second process is the coherent backscattering opposition effect (CBOE) describing constructive interferences of the light scattered coherently backward (Hapke 2002). However, this effect only occurs at very low phase angle, not observable in our experiment.

To make the powder, the original sample was manually powdered but not sieved, leaving a large distribution of grain sizes. The surface texture of the powder can then be seen as rougher than the chip. The SHOE effect is weaker in case of the meteoritic chip as the sample is compact, with little porosity, limiting shadowing of one particle over another. The variation of the BRDF with grain size has already been investigated (Dozier, Schneider, and McGinnis 1981; Pommerol and Schmitt 2008a; Reddy et al. 2015).

Backward scattering has been observed on several small bodies of the Solar System and is often linked to the study of the surface regolith (Belskaya and Shevchenko 2000). Conversely, the study of the bidirectional scattering behavior of a small body, more importantly the strength of the forward scattering and backscattering lobes, can be used to determine the texture, more importantly the presence of dust on the surface.

### 5.2 Bidirectional dependence of the absorption features

The presence of absorption bands in the reflectance spectra of asteroids is, with the spectral slope, commonly used as a criterion to classify small bodies. For example, C-types asteroids exhibits hydration related features around 0.7 µm and 3µm (Takir and Emery 2012). In this section, we present the influence of observation geometry on the detection and depth of absorption features in the bidirectional reflectance spectra of Mukundpura.

For both the chip and powdered samples, the band depths of the 700 nm and 2760 nm features are calculated according to the formula in section 3.2. The results are presented in Figure 13.

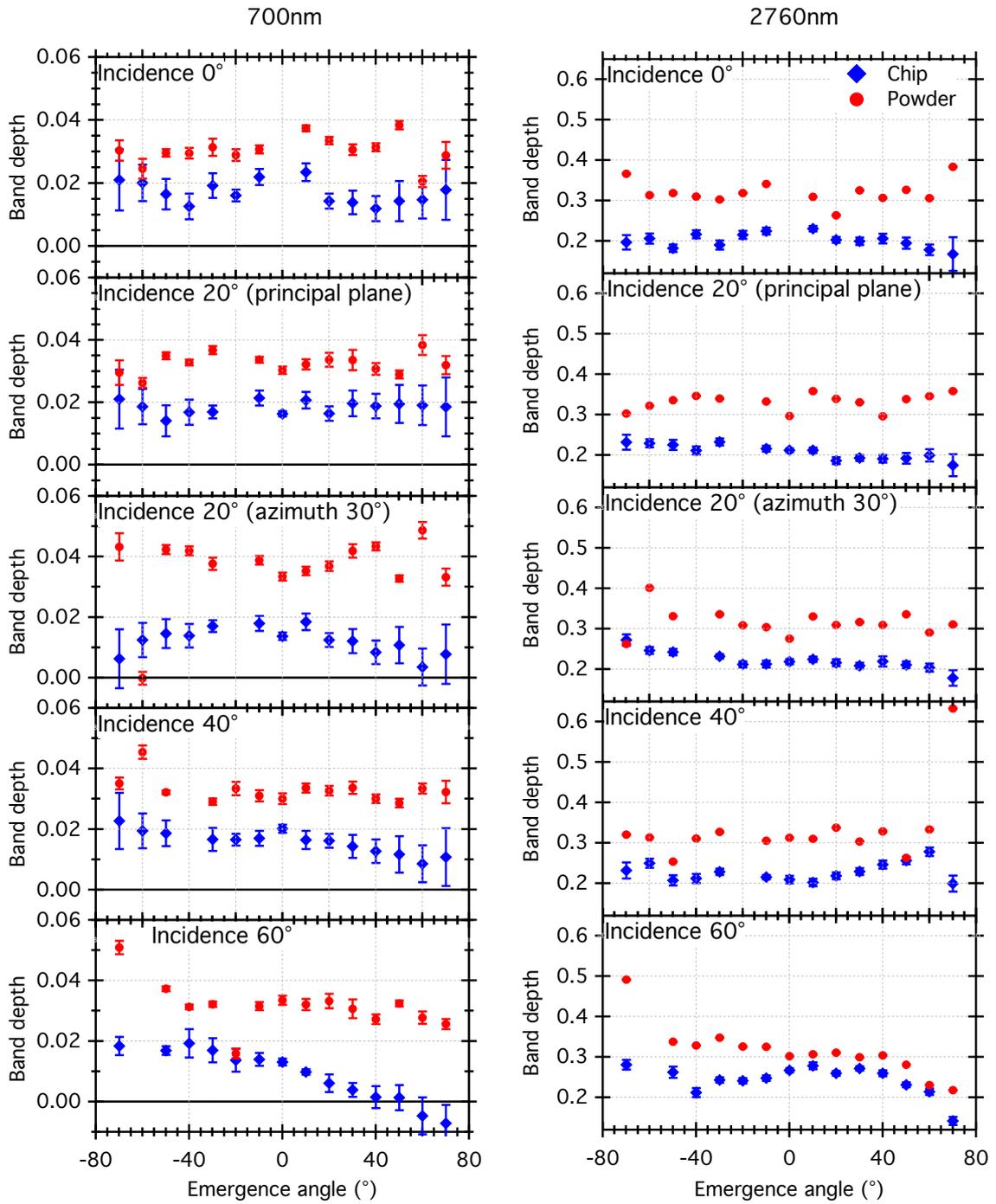

Figure 13: Band depth of the 700nm (left) and the 2760nm (left) absorption features of Mukundpura for the different geometries. Note the different vertical scales.

First, it is important to note that the absorption features detected on the meteoritic chip are generally fainter than for the powdered sample's spectra, in all observation geometries studied. Previous laboratory measurements have shown variation in the band depth of absorption features as a function of geometrical configurations, mostly the decrease of band depth at larger phase angles (Pommerol and Schmitt 2008a; Beck et al. 2012, 2011). This effect is generally explained by an increase of the relative number of photons scattered out of the sample before entering any of the grains. Radiative transfer modeling confirms this hypothesis (Pommerol and Schmitt 2008a).

But several behaviors are observed with our measurements, depending on the incidence angle, not only the phase angle. For illuminations near nadir, the band depth decreases until phase angle around 40°, then increases with the phase angle. In case of grazing illumination, the band depth only decreases with increasing phase angle.

In case of the iron 700nm feature at grazing incidence (60°), the band disappears from the spectra at phase angles larger than 90°, but only on the raw sample. For the powder, the iron band is detectable at all geometrical configurations and the general trend leads us to think that the band will be detectable even at phase angle larger than 120°. The hydration feature at 2760nm is detectable through all our measurements, but begins to fade at 110° phase angle until reaching a depth of 14% at 130°, the largest phase angle studied. According to this trend, the band will also disappear from the spectra, at phase angles greater than 140° or 150° but these are clearly extreme conditions of observation.

The geometry of illumination and observation has an important impact in the depth of the absorption bands. Under all observation geometries, the band depths are stronger for the powder than for the chip. The depth of the 3µm band is variable with the observation geometry but it is always significant. The depth of the 0.7µm band is also variable with observation geometry but can disappear under extreme geometries. Nevertheless, for ground-based observations, mostly obtained at low phase angle, the observation geometry cannot explain the disappearance of the 0.7 µm band.

### 5.3 Geometry dependency of the spectral slope

#### 5.3.1 Phase reddening

Reflectance spectra of most carbonaceous chondrites present a faint red slope, even if blue-sloped spectra can occur under specific conditions (Cloutis et al. 2013). The phase reddening observed on the reflectance spectra of Mukundpura (Figure 11) results from the wavelength dependence of the single-scattering albedo (Gradie and Veverka 1986). This phenomenon was observed on asteroids such as (4) Vesta (Gehrels et al. 1970; Blanco and Catalano 1979), on several Main Belt asteroids (Beth E. Clark et al. 2002), NEAs (Sanchez et al. 2012) and on the Moon (Lane and Irvine 1973; Johnson et al. 2013) and more recently on (101955) Bennu (Binzel et al. 2015). On our spectra, this effect is more important at high incidence angle (at 60°), where observations have been made at the largest phase angles. To have a better overview of the spectral reddening and its dependency with observation geometry Figure 14 presents all spectra after normalization to unity at 1000 nm.

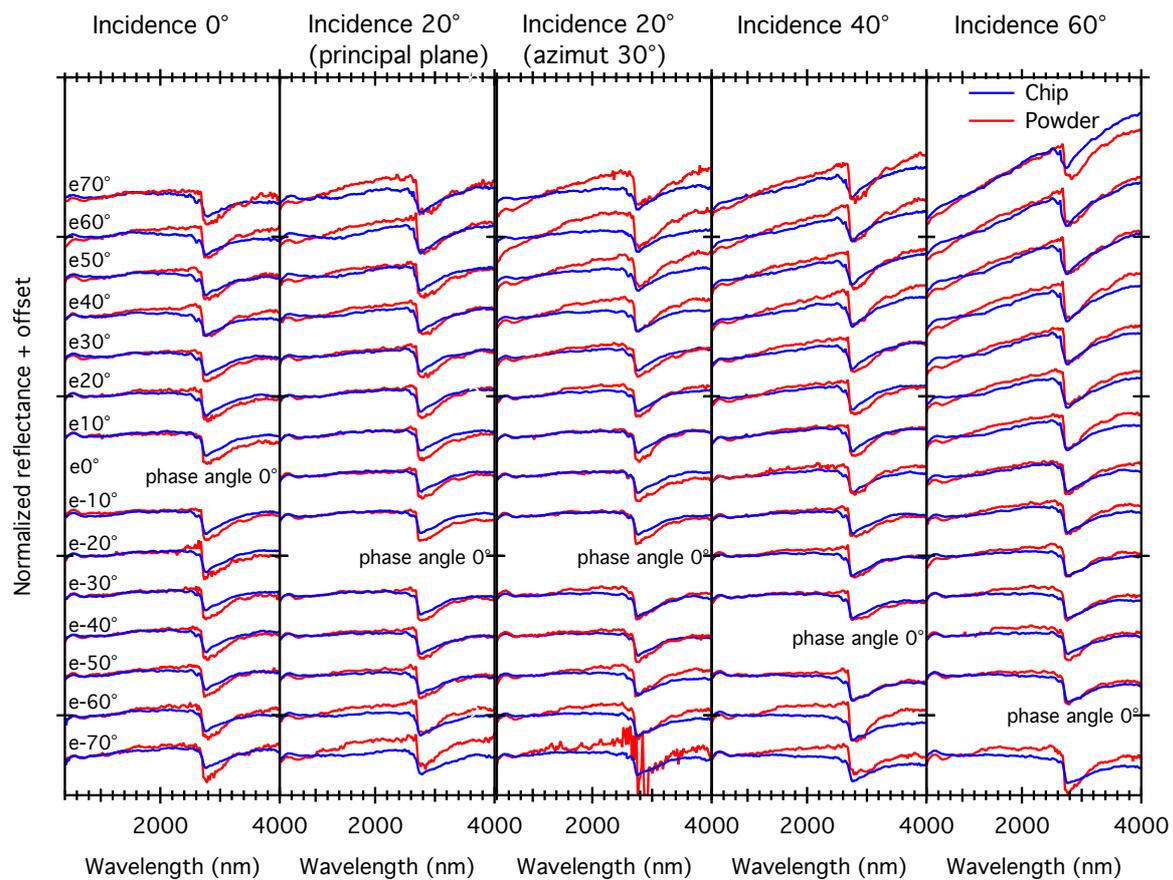

**Figure 14: Reflectance spectra of the Mukundpura meteorite normalized at 1000nm (Blue: chip, Red: powder) taken under different geometrical configurations. Offset for clarity.**

Sanchez et al. (2012) determined the spectral slope as the fitted linear continuum between 0.7 and 1.55µm. This method is commonly used around absorption bands or for measurements on a limited spectral range, but in case of wider spectral ranges, the continuum should not be considered as linear (Parente et al. 2011). The spectral slope can also be calculated as the ratio between the reflectance measured at two separated wavelength, in the visible (Schröder et al. 2014), or between the maximum of reflectance value and the absolute reflectance at 1.5µm (Cloutis et al. 2011) In our case, the spectral range is too wide to adopt a linear approach to the continuum, thus we the spectral slope as a color ratio.

The value of the spectral slope is calculated as the ratio between the reflectance at 2000 nm and the reflectance at 500 nm:

$$Spectral\ slope\ (\mu m^{-1}) = \frac{1}{1.5}\frac{Reflec_{2000\ nm}}{Reflec_{500\ nm}}$$

The variation of the spectral slope with phase angle is displayed on Figure 15, for both sets of spectra.

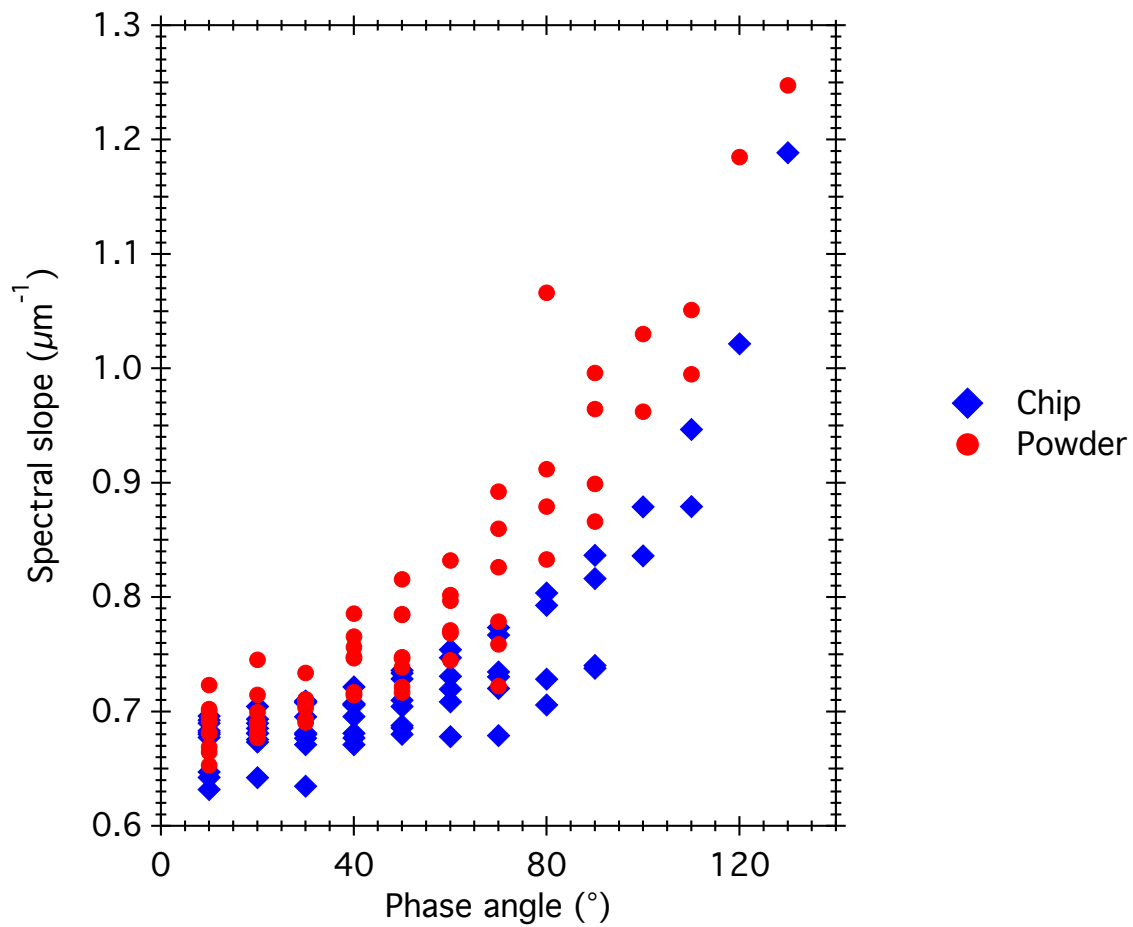

**Figure 15: Effect of the phase angle on the spectral slope of the reflectance spectra (spectral reddening) of the raw Mukundpura meteorite and its powder.**

The reflectance spectra of the powder are redder-sloped than the spectra of the raw meteorite. At low phase angle, the raw meteorite spectra have a slope of 0.6721±0.023 while the powder displays a slope of 0.6861±0.021. This value is averaged over the 10 configurations with a phase angle of 10° and the error corresponds to the standard deviation. At the largest phase angle studied, g = 130°, the slope of the chip spectrum reaches 1.1885±0.0064 compared to 1.2474±0.0090 for the powder spectrum.

Previous investigations pointed out that the surface particle density and texture can radically change the spectral slope from slightly blue to strongly red (Binzel et al. 2015; Britt and Pieters 1988). Cloutis, Hiroi, et al. (2011) showed that a change of porosity applied to the same sample can induce a change in the spectral slope. But the abundance of opaque minerals in the chondrite should not induce significant effects on the spectral reddening (Cloutis, Hudon, et al. 2011). Several laboratory measurements showed that larger grains also induce less red-sloped reflectance spectra (Binzel et al. 2015; Gillis-Davis et al. 2013; Johnson and Fanale 1973; Cloutis, Hudon, et al. 2011; Cloutis et al. 2013).

The variation of slope with the phase angle, i.e. the phase reddening, is pronounced for both raw meteorite and powder. We found that the spectral slope does not vary linearly with the phase angle and that the curve of spectral reddening is different between the meteorite and the powder. In our case, the spectral reddening of the powder is more pronounced at low phase angle but both samples reach roughly the same value at high phase angles. The difference in porosity and granulometry between the two samples induces changes in the spectral slopes. Recently, the analysis of phase reddening has been used to highlight the presence of regolith at the surface of the large boulders of Bennu (DellaGiustina et al. 2019).

### 5.3.2 Bidirectional dependency of the spectral slope

The value of the spectral slope varies with increasing phase angle, as it has been measured for example by the Hayabusa spacecraft on the NEA Itokawa (Abe et al. 2006). Reflectance spectra of the asteroid acquired with a phase angle of 30° showed a spectral slope 6% redder than the spectra at 8° (Abe et al. 2006). This effect was also measured on the NEA (433) Eros, with an increase of the spectral slope reaching 11% between phase angles of 0° and 100° (Clark et al. 2002).

The spectral reddening is calculated for each sample as the ratio between the spectral slope at a given phase angle and the spectral slope at the lowest phase angle g available, so 10°.

$$Spectral\ Reddening = \frac{Spectral\ slope\ _g}{Spectral\ slope\ _{g=10°}}$$

For the two types of surfaces and for each incidence angle, the spectral reddening of the spectra is displayed in Figure 16.

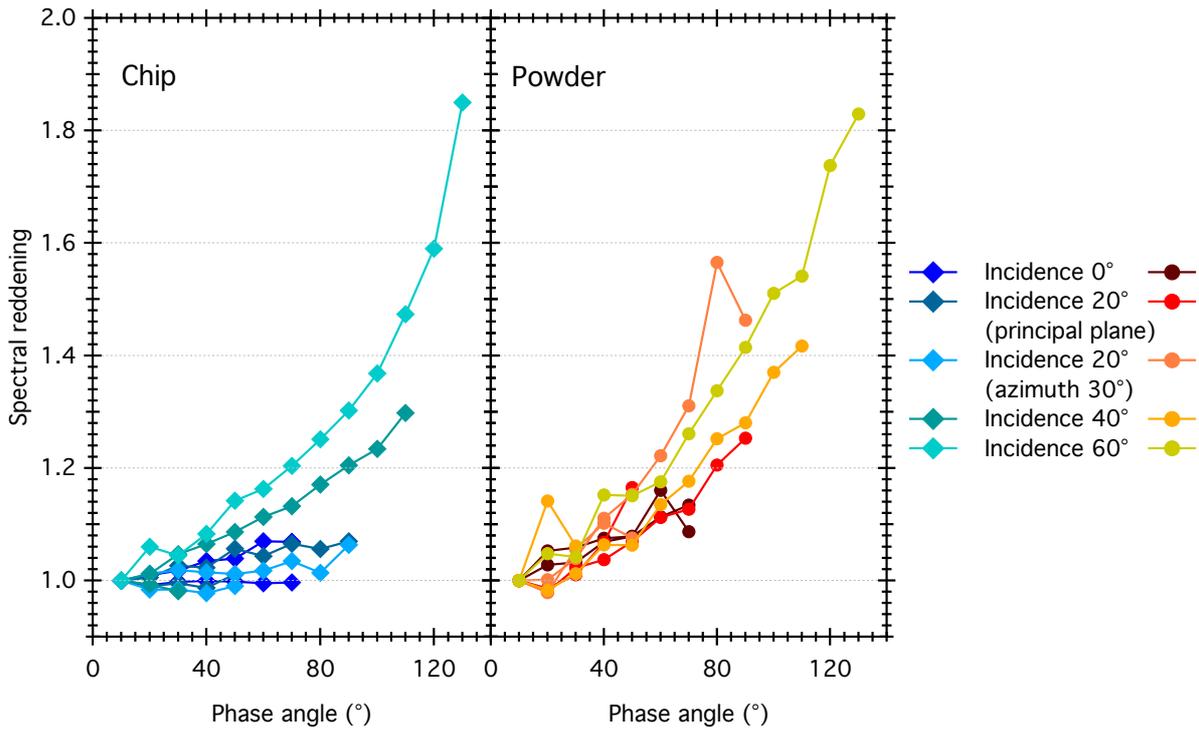

**Figure 16: Spectral reddening of the reflectance spectra of the chip (left) and ground (right) Mukundpura meteorite for each incidence angle.**

Both samples show a dependence of the reddening, not only with phase angle but also with incidence angle. For a phase angle of 70°, the reddening reaches 1.07±0.01 for the meteoritic chip and 1.13±0.02 for the powder measured at nadir incidence, while it reaches 1.20±0.01 for the chip and 1.26±0.01 for the powder for the same phase angle but at incidence 60°. At low incidence angle (≤ 20°) the spectral reddening of the chip always remains weak (< 1.07) even at phase angle as high as 90°, while for the powder the reddening increases steadily even at nadir illumination and is already larger than 1.2 at 80° phase angle. Conversely the spectral reddening seems to depend on the incidence angle at all phase angles for the chip, while for the powder the effect is noticeable only above 60°. Our observations reveal that spectral reddening could be used to detect the presence, or absence of a fine-grained regolith on the surface of a small body by looking at phase reddening at low incidence angle.

Neither the spectral slope nor the spectral reddening show a linear relation with the phase angle, but rather an exponential dependence. MBAs can only be observed with ground-based telescopes at phase angles smaller than 30°, where the reddening is modest (between 1.01 and 1.06 for the powder sample and between 0.98 and 1.05 for the raw meteorite). However, NEAs can be observed over a wider range of phase angles, from opposition to 90° (Sanchez et al. 2012). In this configuration, the reddening of the spectra can reach 1.46 for a dust-covered surface and 1.3 in case of a raw surface.

Particular attention to phase reddening should be given as the variation of spectral slope in asteroid observations is generally mainly attributed to space weathering (irradiation by cosmic ions and micro-meteorites bombardment) (Hapke 2001; Chapman 2004; Clark et al. 2002). Several laboratory studies irradiated meteoritic samples, either with ions or pulsed laser radiations, in order to reproduce the processes occurring on airless bodies. Using the meteorite Epinal, (Strazzulla et al. 2005) showed that ion irradiation increases the spectral slope by a factor of 6 and, comparing the experimental results with spectra of several NEAs, evaluated the timescale for weathering the near-earth asteroids between $10^4$ and $10^6$ years. (Sanchez et al. 2012) measured reflectance spectra on three ordinary chondrites and found that an increase of 30° to 120° of phase angle can mimic the reddening of the reflectance spectrum induced by $0.1 \times 10^6$ to $1.3 \times 10^6$ years of irradiation at 1AU from the Sun. So in case of NEAs, the geometry of observation can induce the same reddening of the spectra as space weathering.

# 6 CONCLUSION

Laboratory reflectance spectroscopy can help understand the specificities of C-complex NEAs observations when compared to C-type MBAs. We have shown that temperature, surface texture, as well as observation geometry (incidence, emergence and azimuth angle) can all have an impact on reflectance spectra. When comparing chips and powdered samples, we found that:

-Both samples display spectral signatures of hydration and thus the absence, or presence, of a fine-grained regolith at the surface cannot explain the lack of hydrated NEAs.

- Dust covered surfaces are expected to be darker than rock covered surfaces because of the apparent homogeneity of the composition of the powder compared to the heterogeneity of the surface of the raw meteorite clearly displaying clasts and inclusions inside a dark matrix.

-BRDF analysis of the two samples shows that the powdered sample presents stronger forward scattering and backscattering lobes compared to the raw meteorite. This is due to the SHOE, increased by the rougher texture of the powder.

-Absorption features appear fainter in the case of the raw sample, due to the higher amount of photons being scattered out of the surface before having entered the sample. Following this result, absorption features should be more difficult to detect on surfaces lacking fine-grained regolith, i.e. rubble pile NEAs.

-At large phase angle, the reflectance spectra of the powdered sample present a steeper spectral slope compared to the measurement on the chip.

-Spectral reddening of the powder appears relatively insensitive to the incident angle while for the raw meteorite at a given phase angle, the reflectance spectra are redder when the illumination approaches grazing incidence.

The surface temperature experienced by NEAs can have an effect on their spectral signatures. The temperature affects the reflectance spectra of a meteorite, mostly by modifying the depth, position and width of the iron features around 700nm and 900nm. Our measurements show that at laboratory timescale, irreversible alterations of the sample can be induced by only one heating-cooling cycle, simulating successively the sunrise and sunset on the surface of the small body. The hydration feature at 3µm becomes fainter but is still detectable on the reflectance spectra. However, our series of measurements after a single thermal cycle at high temperature do not allow us to make any assumption on the possible disappearance of the features after numerous cycles of heating and cooling. Further investigations of these irreversible alterations are currently conducted.

Several effects due to the geometrical configuration of observation have been highlighted by our measurements. First, carbonaceous chondrites do not behave as a lambertian surface in the VNIR. Second the spectral slope and the depths of the absorption features are impacted by the direction of illumination and the emergence angle. Most importantly, our results show that in the case of the raw meteorite the hydration signatures in the visible range disappear from the reflectance spectra at large phase angle. Keeping in mind that NEAs can be observed at larger phase angles compared to MBAs, and that their surfaces seem depleted in fine-grained regolith, this geometry effect could also contribute to the apparent dehydration of NEAs.

**Acknowledgments**

Sandra Potin is supported by Université Grenoble Alpes (UGA) (IRS IDEX/UGA). Pierre Beck acknowledges funding from the H2020 European Research Council (ERC) (SOLARYS ERC-CoG2017_771691). The instrument SHADOWS is founded by the Agence Nationale de la Recherche (ANR) (Labex OSUG@2020, ANR10 LABX56), the Horizon 2020 Framework Programme (H2020) (654208) and the Centre National d'Études Spatiales (CNES).

**References**

Abe, M., Y. Takagi, K. Kitazato, S. Abe, T. Hiroi, F. Vilas, B. E. Clark, et al. 2006. "Near-Infrared Spectral Results of Asteroid Itokawa from the Hayabusa Spacecraft." *Science* 312 (5778): 1334–38. https://doi.org/10.1126/science.1125718.

Beck, P., J. A. Barrat, F. Grisolle, E. Quirico, B. Schmitt, F. Moynier, P. Gillet, and C. Beck. 2011. "NIR Spectral Trends of HED Meteorites: Can We Discriminate between the Magmatic Evolution, Mechanical Mixing and Observation Geometry Effects?" *Icarus* 216 (2): 560–71. https://doi.org/10.1016/j.icarus.2011.09.015.

Beck, P., A. Pommerol, N. Thomas, B. Schmitt, F. Moynier, and J. A. Barrat. 2012. "Photometry of Meteorites." *Icarus* 218 (1): 364–77. https://doi.org/10.1016/j.icarus.2011.12.005.

Beck, P., B. Schmitt, E. A. Cloutis, and P. Vernazza. 2015. "Low-Temperature Reflectance Spectra of Brucite and the Primitive Surface of 1-Ceres?" *Icarus* 257 (June): 471–76.

https://doi.org/10.1016/j.icarus.2015.05.031.

Beck, P, E Quirico, G Montes-hernandez, L Bonal, J Bollard, F Orthous-daunay, K T Howard, et al. 2010. "Hydrous Mineralogy of CM and CI Chondrites from Infrared Spectroscopy and Their Relationship with Low Albedo Asteroids." *Geochimica et Cosmochimica Acta* 74 (16): 4881–92. https://doi.org/10.1016/j.gca.2010.05.020.

Belskaya, I. N., and V. G. Shevchenko. 2000. "Opposition Effect of Asteroids." *Icarus* 147 (1): 94–105. https://doi.org/10.1006/icar.2000.6410.

Benner, Lance A.M., Steven J. Ostro, Christopher Magri, Michael C. Nolan, Ellen S. Howell, Jon D. Giorgini, Raymond F. Jurgens, et al. 2008. "Near-Earth Asteroid Surface Roughness Depends on Compositional Class." *Icarus* 198 (2): 294–304. https://doi.org/10.1016/j.icarus.2008.06.010.

Binzel, Richard P., Francesca E. DeMeo, Brian J. Burt, Edward A. Cloutis, Ben Rozitis, Thomas H. Burbine, Humberto Campins, et al. 2015. "Spectral Slope Variations for OSIRIS-REx Target Asteroid (101955) Bennu: Possible Evidence for a Fine-Grained Regolith Equatorial Ridge." *Icarus* 256: 22–29. https://doi.org/10.1016/j.icarus.2015.04.011.

Blanco, C, and S Catalano. 1979. "UBV Photometry of Vesta." *Icarus* 40: 350–63. https://doi.org/https://doi.org/10.1016/0019-1035(79)90028-9.

Bland, P. A., G. S. Collins, T. M. Davison, N. M. Abreu, F. J. Ciesla, A. R. Muxworthy, and J. Moore. 2014. "Pressure-Temperature Evolution of Primordial Solar System Solids during Impact-Induced Compaction." *Nature Communications* 5: 1–13. https://doi.org/10.1038/ncomms6451.

Brearley, Adrian J. 2006. "The Action of Water." In *Meteorites and the Early Solar System II*, 587–624. https://doi.org/10.1039/AN9214600270.

Brissaud, Olivier, Bernard Schmitt, Nicolas Bonnefoy, Sylvain Douté, Patrick Rabou, Will Grundy, and Michel Fily. 2004. "Spectrogonio Radiometer for the Study of the Bidirectional Reflectance and Polarization Functions of Planetary Surfaces. 1. Design and Tests." *Applied Optics* 43 (9): 1926–37. https://doi.org/10.1364/AO.43.001926.

Britt, D. T., and C. M. Pieters. 1988. "Bidirectional Reflectance Properties of Iron-Nickel Meteorites." In *18th LPSC*, 503–12.

Burns, R.G. 1993. *Mineralogical Applications of Crystal Field Theory, Second Edition*. Cambdrige University Press.

Carry, B., E. Solano, S. Eggl, and F. E. DeMeo. 2016. "Spectral Properties of Near-Earth and Mars-Crossing Asteroids Using Sloan Photometry." *Icarus* 268: 340–54. https://doi.org/10.1016/j.icarus.2015.12.047.

Chapman, Clark R. 2004. "Space Weathering of Asteroid Surfaces." *Annual Review of Earth and Planetary Sciences* 32 (1): 539–67. https://doi.org/10.1146/annurev.earth.32.101802.120453.

Clark, Beth E., P. Helfenstein, J. F. Bell, C. Peterson, J. Veverka, N. I. Izenberg, D. Domingue, D. Wellnitz, and Lucy McFadden. 2002. "NEAR Infrared Spectrometer Photometry of Asteroid 433 Eros." *Icarus* 155 (1): 189–204. https://doi.org/10.1006/icar.2001.6748.

Clark, Beth Ellen, Bruce Hapke, Carle Pieters, and Daniel Britt. 2002. "Asteroid Space Weathering and Regolith Evolution." In *Asteroids III*, 585–99. https://doi.org/10.1016/0273-1177(91)90551-T.

Cloutis, E. A., T. Hiroi, M. J. Gaffey, C. M.O.D. Alexander, and P. Mann. 2011. "Spectral Reflectance Properties of Carbonaceous Chondrites: 1. CI Chondrites." *Icarus* 212 (1): 180–209. https://doi.org/10.1016/j.icarus.2010.12.009.

Cloutis, E. A., P. Hudon, T. Hiroi, M. J. Gaffey, and P. Mann. 2011. "Spectral Reflectance Properties of Carbonaceous Chondrites: 2. CM Chondrites." *Icarus* 216 (1): 309–46. https://doi.org/10.1016/j.icarus.2011.09.009.

Cloutis, Edward A., P. Hudon, Takahiro Hiroi, Michael J. Gaffey, Paul Mann, Conel M. O'D. Alexander, James F. Bell III, and Beth Ellen Clark. 2013. "Possible Causes of Blue Slopes (~0.5-2.5 Mm) in Carbonaceous Chondrite Spectra." *44th Lunar and Planetary Science Conference*, 1550.

Cloutis, Edward A., Kaitlyn A. McCormack, James F. Bell, Amanda R. Hendrix, Daniel T. Bailey, Michael A. Craig, Stanley A. Mertzman, Mark S. Robinson, and Miriam A. Riner. 2008. "Ultraviolet Spectral Reflectance Properties of Common Planetary Minerals." *Icarus* 197 (1): 321–47. https://doi.org/10.1016/j.icarus.2008.04.018.

Cloutis, Edward A., Valerie B. Pietrasz, Cain Kiddell, Matthew R.M. Izawa, Pierre Vernazza, Thomas H. Burbine, Francesca DeMeo, et al. 2018. "Spectral Reflectance 'Deconstruction' of the Murchison CM2 Carbonaceous Chondrite and Implications for Spectroscopic Investigations of Dark Asteroids." *Icarus* 305: 203–24. https://doi.org/10.1016/j.icarus.2018.01.015.

Delbo, Marco, and Patrick Michel. 2011. "Temperature History and Dynamical Evolution of (101955) 1999 RQ 36: A Potential Target for Sample Return from a Primitive Asteroid." *Astrophysical Journal Letters* 728 (2 PART II): 1–5. https://doi.org/10.1088/2041-8205/728/2/L42.

DellaGiustina, D. N., J. P. Emery, D. R. Golish, B. Rozitis, C. A. Bennett, K. N. Burke, R. L. Ballouz, et al. 2019. "Properties of Rubble-Pile Asteroid (101955) Bennu from OSIRIS-REx Imaging and Thermal Analysis." *Nature Astronomy* 3 (4): 341–51. https://doi.org/10.1038/s41550-019-0731-1.

Dozier, Jeff, Stanley R. Schneider, and David F. McGinnis. 1981. "Effect of Grain Size and Snowpack Water Equivalence on Visible and Near-infrared Satellite Observations of Snow." *Water Resources Research* 17 (4): 1213–21. https://doi.org/10.1029/WR017i004p01213.

Dunn, Tasha L., Thomas H. Burbine, William F. Bottke, and John P. Clark. 2013. "Mineralogies and Source Regions of Near-Earth Asteroids." *Icarus* 222 (1): 273–82. https://doi.org/10.1016/j.icarus.2012.11.007.

Fornasier, S., C. Lantz, M. A. Barucci, and M. Lazzarin. 2014. "Aqueous Alteration on Main

Belt Primitive Asteroids: Results from Visible Spectroscopy." *Icarus* 233: 163–78. https://doi.org/10.1016/j.icarus.2014.01.040.

Fujiwara, A, J Kawaguchi, D K Yeomans, M Abe, T Mukai, T Okada, J Saito, et al. 2006. "The Rubble-Pile Asteroid Itokawa as Observed by Hayabusa." *Science* 312 (5778): 1330–34. https://doi.org/10.1126/science.1125841.

Garenne, A., P. Beck, G. Montes-Hernandez, R. Chiriac, F. Toche, E. Quirico, L. Bonal, and B. Schmitt. 2014. "The Abundance and Stability of 'Water' in Type 1 and 2 Carbonaceous Chondrites (CI, CM and CR)." *Geochimica et Cosmochimica Acta* 137: 93–112. https://doi.org/10.1016/j.gca.2014.03.034.

Gehrels, T, E Roemer, R.C. Taylor, and B.H Zllner. 1970. "Minor Planets and Related Objects. IV. Asteroid (1566) Icarus." *The Astronomical Journal* 75 (2): 186–95.

Ghosh, Weidenschilling, McSween, and Rubin. 2006. "Asteroidal Heating and Thermal Stratification of the Asteroid Belt." In *Meteorites and the Early Solar System II*, 555–66. https://doi.org/10.1152/ajpgi.00524.2002.

Gillis-Davis, Jeffrey J., P.G. Lucey, J. P. Bradley, H. Ishii, and H. C. Connolly. 2013. "LASER SPACE WEATHERING OF ALLENDE METEORITE." *44th Lunar and Planetary Science Conference*, 2494. https://doi.org/10.1029/2005JE002544.

Gradie, J, and J. Veverka. 1986. "The Wavelength Dependence of Phase Coefficients." *Icarus* 66: 455–67.

Grimm, R.E., and H.Y. Mcsween. 1993. "Heliocentric Zoning of the Asteroid Belt by Aluminum-26." *Science* 259: 653–55.

Grimm, Robert E., and Harry Y. Mcsween. 1989. "Water and the Thermal Evolution of Carbonaceous Chondrite Parent Bodies." *Icarus* 82 (2): 244–80. https://doi.org/10.1016/0019-1035(89)90038-9.

Grisolle, F, B Schmitt, P Beck, S Philippe, and O Brissaud. 2014. "Experimental Simulation of the Condensation and Metamorphism of Seasonal CO2 Condensates under Martian Conditions." In *EPSC Abstracts*, 9:EPSC2014-635.

Hamilton, V. E., A A Simon, P R Christensen, D. Reuter, B. E. Clark, M A Barucci, N.E. Bowles, et al. 2019. "Evidence for Widespread Hydrated Minerals on Asteroid (101955) Bennu." *Nature Astronomy* 3 (101955): 332–40. https://doi.org/https://doi.org/10.1038/s41550-019-0722-2.

Hapke, Bruce. 1981. "Bidirectional Reflectance Spectroscopy: 1. Theory." *Journal of Geophysical Research: Solid Earth* 86 (B4): 3039–54. https://doi.org/10.1029/JB086iB04p03039.

———. 1986. "Bidirectional Reflectance Spectroscopy. 4. The Extinction Coefficient and the Opposition Effect." *Icarus* 67 (2): 264–80. https://doi.org/10.1016/0019-1035(86)90108-9.

———. 2001. "Space Weathering from Mercury to the Asteroid Belt." *Journal of Geophysical Research* 106: 10039–73. https://doi.org/doi:10.1029/2000JE001338.

———. 2002. "Bidirectional Reflectance Spectroscopy. 5. The Coherent Backscatter Opposition Effect and Anisotropic Scattering." *Icarus* 157 (2): 523–34. https://doi.org/10.1006/icar.2002.6853.

———. 2008. "Bidirectional Reflectance Spectroscopy. 6. Effects of Porosity." *Icarus* 195 (2): 918–26. https://doi.org/10.1016/j.icarus.2008.01.003.

Hasegawa, Sunao, Thomas G. Müller, Kyoko Kawakami, Toshihiro Kasuga, Takehiko Wada, Yoshifusa Ita, Naruhisa Takato, Hiroshi Terada, Takuya Fujiyoshi, and Masanao Abe. 2008. "Albedo, Size, and Surface Characteristics of Hayabusa-2 Sample-Return Target 162173 1999 JU3 from AKARI and Subaru Observations." *Publications of the Astronomical Society of Japan* 60 (SP2): 399–405. https://doi.org/".

Hinrichs, John L., and Paul G. Lucey. 2002. "Temperature-Dependent near-Infrared Spectral Properties of Minerals, Meteorites, and Lunar Soil." *Icarus* 155 (1): 169–80. https://doi.org/10.1006/icar.2001.6754.

Johnson, Jeffrey R., Michael K. Shepard, William M. Grundy, David A. Paige, and Emily J. Foote. 2013. "Spectrogoniometry and Modeling of Martian and Lunar Analog Samples and Apollo Soils." *Icarus* 223 (1): 383–406. https://doi.org/10.1016/j.icarus.2012.12.004.

Johnson, Torrence V, and Fraser P Fanale. 1973. "Optical Properties of Carbonaceous Chondrites and Their Relationship to Asteroids." *Journal of Geophysical Research* 78 (35): 8507–18. https://doi.org/10.1029/JB078i035p08507.

Kitazato, K., R. E. Milliken, T. Iwata, M. Abe, M. Ohtake, S. Matsuura, T. Arai, et al. 2019. "The Surface Composition of Asteroid 162173 Ryugu from Hayabusa2 Near-Infrared Spectroscopy." *Science*, 10.1126/science.aav7432. https://doi.org/10.1126/science.aav7432.

Lane, Adair P, and William M Irvine. 1973. "Monochromatic Phase Curves and Albedos for the Lunar Disk." *The Astronomical Journal* 78 (3): 267–77.

Mahan, Brandon, Frédéric Moynier, Pierre Beck, Emily A. Pringle, and Julien Siebert. 2018. "A History of Violence: Insights into Post-Accretionary Heating in Carbonaceous Chondrites from Volatile Element Abundances, Zn Isotopes and Water Contents." *Geochimica et Cosmochimica Acta* 220: 19–35. https://doi.org/10.1016/j.gca.2017.09.027.

Marchi, S., M. Delbo', A. Morbidelli, P. Paolicchi, and M. Lazzarin. 2009. "Heating of Near-Earth Objects and Meteoroids Due to Close Approaches to the Sun." *Monthly Notices of the Royal Astronomical Society* 400 (1): 147–53. https://doi.org/10.1111/j.1365-2966.2009.15459.x.

Michel, Patrick, Willy Benz, and Derek C. Richardson. 2003. "Disruption of Fragmented Parent Bodies as the Origin of Asteroid Families." *Nature* 421 (6923): 608–11. https://doi.org/10.1038/nature01364.

———. 2004. "Catastrophic Disruption of Asteroids and Family Formation: A Review of Numerical Simulations Including Both Fragmentation and Gravitational Reaccumulations." *Planetary and Space Science* 52 (12): 1109–17.

https://doi.org/10.1016/j.pss.2004.07.008.

Nakamura, T. 2005. "Post-Hydration Thermal Metamorphism of Carbonaceous Chondrites." *Journal of Mineralogical and Petrological Sciences* 100: 260–72.

Nakamura, Tomoki. 2006. "Yamato 793321 CM Chondrite: Dehydrated Regolith Material of a Hydrous Asteroid." *Earth and Planetary Science Letters* 242 (1–2): 26–38. https://doi.org/10.1016/j.epsl.2005.11.040.

Parente, Mario, Heather D. Makarewicz, and Janice L. Bishop. 2011. "Decomposition of Mineral Absorption Bands Using Nonlinear Least Squares Curve Fitting: Application to Martian Meteorites and CRISM Data." *Planetary and Space Science* 59 (5–6): 423–42. https://doi.org/10.1016/j.pss.2011.01.009.

Pommerol, Antoine, and Bernard Schmitt. 2008a. "Strength of the H2O Near-Infrared Absorption Bands in Hydrated Minerals: Effects of Measurement Geometry." *Journal of Geophysical Research E: Planets* 113 (12): E12008. https://doi.org/10.1029/2008JE003197.

———. 2008b. "Strength of the H2O Near-Infrared Absorption Bands in Hydrated Minerals: Effects of Particle Size and Correlation with Albedo." *Journal of Geophysical Research* 113 (E10): E10009. https://doi.org/10.1029/2007JE003069.

Pommerol, Antoine, Bernard Schmitt, Pierre Beck, and Olivier Brissaud. 2009. "Water Sorption on Martian Regolith Analogs: Thermodynamics and near-Infrared Reflectance Spectroscopy." *Icarus* 204 (1): 114–36.

https://doi.org/10.1016/j.icarus.2009.06.013.

Potin, Sandra, Olivier Brissaud, Pierre Beck, Bernard Schmitt, Yves Magnard, Jean-Jacques Correia, Patrick RAbou, and Laurent Jocou. 2018. "SHADOWS : A Spectro-Gonio Radiometer for Bidirectional Reflectance Studies of Dark Meteorites and Terrestrial Analogs . Design , Calibrations , and Performances on Challenging Surfaces." *Applied Optics* 57 (29): 8279–8296. https://doi.org/https://doi.org/10.1364/AO.57.008279.

Ray, Dwijesh, and Anil D. Shukla. 2018. "The Mukundpura Meteorite, a New Fall of CM Chondrite." *Planetary and Space Science* 151: 149–54. https://doi.org/10.1016/j.pss.2017.11.005.

Reddy, V., T. L. Dunn, C. A. Thomas, N. A. Moskovitz, and T. H. Burbine. 2015. "Mineralogy and Surface Composition of Asteroids." *Asteroids IV*, no. 2867. https://doi.org/10.2458/azu_uapress_9780816532131-ch003.

Rivkin, A S, E S Howell, F Vilas, and L A Lebofsky. 2002. "Hydrated Minerals on Asteroids: The Astronomical Record." In *Asteroids III*, 235–53. https://doi.org/10.1086/300495.

Rivkin, Andrew S., and F. E. DeMeo. 2018. "How Many Hydrated NEOs Are There?" *Journal of Geophysical Research: Planets* 123 (December): (accepted). https://doi.org/10.1029/2018JE005584.

Sanchez, Juan A., Vishnu Reddy, Andreas Nathues, Edward A. Cloutis, Paul Mann, and

Harald Hiesinger. 2012. "Phase Reddening on Near-Earth Asteroids: Implications for Mineralogical Analysis, Space Weathering and Taxonomic Classification." *Icarus* 220 (1): 36–50. https://doi.org/10.1016/j.icarus.2012.04.008.

Schorghofer, N. 2008. "Lifetime of Ice on Main Belt Asteroids." *Astrophysical Journal* 682 (1): 697–705. https://doi.org/10.1086/588633.

Schröder, S. E., Ye Grynko, A. Pommerol, H. U. Keller, N. Thomas, and T. L. Roush. 2014. "Laboratory Observations and Simulations of Phase Reddening." *Icarus* 239: 201–16. https://doi.org/10.1016/j.icarus.2014.06.010.

Shepard, Michael K., and Paul Helfenstein. 2011. "A Laboratory Study of the Bidirectional Reflectance from Particulate Samples." *Icarus* 215 (2): 526–33. https://doi.org/10.1016/j.icarus.2011.07.033.

Smith, G. 1977. "Low.Temperature Optical Studies of Metal-Metal Charge.Transfer Transitions in Various Minerals." *Canadian Mineralogist* 15: 500–507. http://rruff.info/doclib/cm/vol15/CM15_500.pdf.

Strazzulla, Gianni, E. Dotto, R. Binzel, R. Brunetto, M. A. Barucci, A. Blanco, and V. Orofino. 2005. "Spectral Alteration of the Meteorite Epinal (H5) Induced by Heavy Ion Irradiation: A Simulation of Space Weathering Effects on near-Earth Asteroids." *Icarus* 174 (1): 31–35. https://doi.org/10.1016/j.icarus.2004.09.013.

Stuart, Joseph Scott, and Richard P. Binzel. 2004. "Bias-Corrected Population, Size Distribution, and Impact Hazard for the near-Earth Objects." *Icarus* 170 (2): 295–

311. https://doi.org/10.1016/j.icarus.2004.03.018.

Sunshine, Jessica M, Carle M Pieters, and Stephen F Pratt. 1990. "Deconvolution of Mineral Absorption Bands : An Improved Approach." *Journal of Geophysical Research* 95: 6955–66.

Takir, Driss, and Joshua P. Emery. 2012. "Outer Main Belt Asteroids: Identification and Distribution of Four 3-Mm Spectral Groups." *Icarus* 219 (2): 641–54. https://doi.org/10.1016/j.icarus.2012.02.022.

Takir, Driss, Joshua P. Emery, Harry Y. Mcsween, Charles A. Hibbitts, Roger N. Clark, Neil Pearson, and Alian Wang. 2013. "Nature and Degree of Aqueous Alteration in CM and CI Carbonaceous Chondrites." *Meteoritics and Planetary Science* 48 (9): 1618–37. https://doi.org/10.1111/maps.12171.

Tomeoka, Kazushige, Harry Y Mcsween, and Peter R Buseck. 1989. "Mineralogical Alteration of CM Carbonaceous Chondrites: A Review." In *Proc. NIPR Symp. Antarct. Meteorites*, 2:221–34.

Tripathi, R. P., Ambesh Dixit, and N. Bhandari. 2018. "Characterization of Mukundpura Carbonaceous Chondrite." *Current Science* 114 (1): 214–17. https://doi.org/10.18520/cs/v114/i01/214-217.

Usui, Fumihiko, Sunao Hasegawa, Takafumi Ootsubo, and Takashi Onaka. 2019. "AKARI/IRC Near-Infrared Asteroid Spectroscopic Survey: AcuA-Spec." *Publ. Astron. Soc. Japan* 71 (0): 1–41. https://doi.org/10.1093/pasj/xxx000.

Veverka, J, P Thomas, A Harch, B Clark, J F Bell Iii, M Malin, L A Mcfadden, S Murchie, and S E Hawkins Iii. 1997. "NEAR ' s Flyby of 253 Mathilde : Images of a C Asteroid." *Science* 278 (December 1997): 2109–14.

Vilas, Faith. 1994. "A Cheaper, Faster, Better Way to Detect Water of Hydratation on Solar System Bodies." *Icarus* 111: 456–67.

———. 2008. "Spectral Characteristics of Hayabusa 2 Near-Earth Asteroid Targets 162173 1999 JU3 and 2001 QC34." *Astronomical Journal* 135 (4): 1101–5. https://doi.org/10.1088/0004-6256/135/4/1101.